\documentclass{aa}
\usepackage[utf8]{inputenc}
\usepackage[varg]{txfonts}
\usepackage{physics}
\usepackage{gensymb}
\usepackage{nameref}
\usepackage{listings}
\usepackage{graphicx}
\usepackage{multirow}
\usepackage{breqn}
\usepackage{textcomp}
\begin{document}
\title{Kinematic constraints on the ages and kick velocities of Galactic neutron star binaries}

\author{
P. Disberg \inst{1}
\and
N. Gaspari\inst{1}
\and
A. J. Levan\inst{1,2}
}

\institute{Department of Astrophysics/IMAPP, Radboud University, P.O. Box 9010, 6500 GL Nijmegen, The Netherlands\\
e-mail: paul.disberg@gmail.com
\and
Department of Physics, University of Warwick, Coventry CV4 7AL, UK}

\date{\today}

\abstract{The systems creating binary neutron stars (BNSs) experience systemic kicks when one of the components goes supernova. The combined magnitude of these kicks is still a topic of debate, and has implications for the eventual location of the transient resulting from the merger of the binary. For example, the offsets of short-duration gamma-ray bursts (SGRBs) resulting from BNS mergers depend on the BNS kicks.}{We investigated Galactic BNSs, and traced their motion through the Galaxy. This enabled us to estimate their kinematic ages and construct a BNS kick distribution, based on their Galactic trajectories.}{We used the pulsar periods and their derivatives to estimate the characteristic spin-down ages of the binaries. Moreover, we used a Monte Carlo estimation of their present-day velocity vector in order to trace back their trajectory and estimate their kinematic ages. These trajectories, in turn, were used to determine the eccentricity of their Galactic orbit. Based on simulations of kicked objects in the Galactic potential, we investigated the relationship between this eccentricity and kick velocity, in order to constrain the kicks imparted to the binaries at birth.}{We find that the Galactic BNSs are likely older than $\sim40$ Myr, which means their current (scalar) galactocentric speeds are not representative of their initial kicks. However, we find a close relationship between the eccentricity of a Galactic trajectory and the experienced kick. Using this relation, we constrained the kicks of the Galactic BNSs, depending on the kind of isotropy assumed in estimating their velocity vectors. These kick velocities are well-described by a log-normal distribution peaking around $\sim40-50$ km/s, and coincide with the peculiar velocities of the binaries at their last disc crossing.}{We conclude that BNSs receive kicks following a distribution that peaks at kick velocities lower than found in isolated pulsars. However, we find no tension between this distribution and literature on SGRB offsets.}
\maketitle

\section{Introduction}
\label{sec1}
Neutron star binaries -- and in particular their mergers -- can provide insight into a variety of astrophysical phenomena. For example, mergers of these binary neutron stars (BNSs), consisting of two neutron stars (NSs) orbiting each other, have produced gravitational waves that have been detected \citep[e.g.][]{Abbott_2017} and are used to gain insight into the structure of NSs \citep[e.g.][]{Radice_2018}. Moreover, BNS mergers are thought to produce short-duration gamma-ray bursts \citep[SGRBs;][]{Eichler_1989,Narayan_1992}, which is why their rates and locations have been topics of research \citep[for a review see][]{Berger_2014}. Furthermore, the mergers of such binaries are a prime source of heavy element ($r-$process) enrichment throughout the Universe, and so far the only site for which direct evidence of heavy element production has been found \citep[][for a review see \citeauthor{Nakar_2020} \citeyear{Nakar_2020}]{Berger_2013,Tanvir_2013,Arcavi_2017,Kasen_2017,Metzger_2017,Pian_2017,Smartt_2017,Tanvir_2017,Villar_2017,Rastinejad_2022,Troja_2022,Levan_2023}. Understanding both the details of the mergers (e.g. component masses, ejecta masses, composition) and also the locations of the mergers relative to host galaxies is then critical to determining their role in cosmic chemical evolution.\\
\indent After all, the merger locations of BNS with respect to their host galaxies are determined by the velocity kicks these systems receive on formation \citep[e.g.,][]{Fryer_1999,Bloom_1999,Bulik_1999,Belczynski_2002,Perna_2002,Voss_2003,Belczynski_2006,Church_2011,Abbott_2017host,Zevin_2020}. That is, both NSs are formed in a supernova, and because of this receive a natal kick, whose magnitude has been estimated by analysing the velocities of young pulsars \citep{Lyne_1994,Hobbs_2005,Faucher_2006,Arzoumanian_2002,Verbunt_2017,Igoshev_2020,Igoshev_2021}. The total systemic kick the binary receives is then a combination of these two natal kicks and a Blaauw kick due to mass loss at each supernova \citep{Blaauw_1961,Hills_1983,VandenHeuvel_2000,Zhang_2013,Tauris_2017,Andrews_2019b,Zhao_2023}. The magnitude of the BNSs' systemic kicks has been debated: high kicks can, for instance, explain SGRBs that appear to be host-less \citep{Berger_2010,Fong_2013,Tunnicliffe_2014,O'Connor_2022,Fong_2022}. In particular, while generically large kicks like those seen in pulsars can recreate the observed offsets to short-GRBs reasonably well \citep{Church_2011}, \citet{Behroozi_2014} find that the galaxy-offset distribution of SGRBs, as found by \cite{Fong_2013}, can be explained by a SGRB progenitor kick distribution where only $19\%$ of objects are kicked with velocities $>150$ km/s. Indeed, large kicks may not be necessary to explain SGRBs \citep{Perets_2021} and Galactic BNSs appear also to have experienced small kicks, as suggested by their orbital parameters \citep{Beniamini_2016} and their small peculiar velocities \citep{Gaspari_2024}.\\
\indent We are interested in investigating whether we can constrain the systemic kicks of the Galactic BNSs, for instance by analysing their current speeds. \citet{Disberg_2024} show, however, that kicked objects near the Solar System that are older than a few tens of Myr will obtain galactocentric speeds that are not representative of their kicks, due to their motion within the Galactic potential. For this reason we are interested in estimating the ages of the Galactic BNSs, and aim to give an overview of their spin-down ages, based on braking in rotational speed of the NS in the binary that is observed as pulsar, and their kinematic ages, based on their distance from the Galactic plane. For some of the Galactic BNSs there already exist estimates of the spin-down ages \citep[e.g.][]{Lorimer_2005,Kargaltsev_2006,Andrews_2019} and the kinematic ages \citep[e.g.][]{Arzoumanian_1999,Wex_2000,Willems_2004,Willems_2006}, but we aim to give a complete overview for the Galactic BNSs systems that have known location, proper motion and no association with a globular cluster \citep[as e.g.\ listed by][]{Ding_2024}.\\
\indent In order to investigate the kicks of the BNSs, we analysed the magnitude of their current galactocentric velocities \citep[Monte Carlo estimated following][]{Gaspari_2024}, taking into account the estimates for their ages and the deceleration found by \citet{Disberg_2024}. However, our estimates contain the complete velocity vectors, not just their magnitude, which provides additional information that can potentially be used to constrain their kicks. \citet{Atri_2019}, for instance, use estimated velocity vectors to trace back trajectories of black hole X-ray binaries to the Galactic plane, where they are thought to be formed, and determine the peculiar velocity at the disc crossing as a potential kick velocity \citep[cf.][who apply this method to binaries with a NS component]{O'Doherty_2023}.\\
\indent Likewise, we used the estimated velocity vectors in order to trace back the trajectories of the Galactic BNSs, but we are interested in using the shape of the complete trajectory in order to estimate their kicks. That is, \citet{Disberg_2024} find that the initially circular Galactic orbits of kicked objects are disturbed by the kick, resulting in more eccentric Galactic orbits \citep[as also noted by e.g.][]{Hoang_2022}. We are interested in investigating whether we can use this fact to constrain kicks based on the eccentricity of the Galactic orbits of the BNS systems.\\
\indent This paper is structured as follows. In Sect.\ \ref{sec2} we investigate the ages of the Galactic BNSs. Then, in Sect.\ \ref{sec3}, we analyse the current speeds of the BNSs and use the work of \citet{Disberg_2024} to show that these harbour little information about their kicks. As an alternative method to constrain kicks, we show in Sect.\ \ref{sec4} that the eccentricity of a Galactic orbit is a better indication of kick velocity, and estimate these eccentricities for the Galactic BNSs. In Sect.\ \ref{sec5}, we determine the relationship between kick velocity and eccentricity of the Galactic orbit, and use this relationship to kinematically constrain the kicks of the BNSs. Finally, in Sect.\ \ref{sec6}, we summarise our findings and their implications.
\section{Ages}
\label{sec2}
We are interested in investigating the ages of the Galactic BNSs, since a young population of kicked objects might look different from an old population of kicked objects, due to their motion through the Galaxy \citep{Disberg_2024}. In order to do this, we used two quantities that can tell us something about their ages: spin-down ages (Sect.\ \ref{sec2.1}) and kinematic ages (Sect.\ \ref{sec2.2}).
\begin{table*}
\centering
\caption{Spin properties of Galactic BNS pulsars: the period ($P$), derivative of the period ($\dot{P}$), frequency ($\nu$), derivative of the frequency ($\dot{\nu}$), and the references (Ref.) for the listed values.}
\label{tab1}
\begin{tabular}{lccccc}
		\hline
		\hline\\[-10pt]
		BNS pulsar                   & $P$                   & $\dot{P}$         & $\nu$                 & $\dot{\nu}$            & Ref. \\
				             & [$10^{-3}\ s$]        & [$10^{-18}\ s s^{-1}$] & [$s^{-1}$]       & [$10^{-15}\ s^{-2}$] &  \\ \hline\\[-10pt]
		J0737--3039A\tablefootmark{a} & (...)                 & (...)             & $44.05406864196281(17)$  & $-3.4158071(11)$         &  1 \\
		J0737--3039B\tablefootmark{a} & (...)                 & (...)             & $0.36056035506(1)$    & $-0.116(1)$          &  2    \\
		B1913+16                     & (...)                 & (...)             & $16.940537785677(3)$  & $-2.4733(1)$         &  3 \\
		J1913+1102                   & $24.2850068680286(19)$& $0.15672(7)$      & (...)                 & (...)                &  4   \\
		J0509+3801                   & $76.5413487220(1)$  & $7.931(2)$        & (...)                 & (...)                &  5   \\
		J1756--2251                   & (...)                 & (...)             & $35.1350727145469(6)$ & $-1.256079(3)$       &  6    \\
		B1534+12                     & (...) & (...)     & $26.38213277689397(11)$                & $-1.686097(2)$                &  7 \\
		J1829+2456                   & (...)  		     & (...)             & $24.384401411040(6)
$  & $-0.029395(14)$      &  8   \\
		J1411+2551                   & $62.452895517590(2)$  & $0.0956(51)$      & (...)                 & (...)                &  9    \\
		J0453+1559                   & $45.7818168729515(33)$  & $0.18612(8)$      & (...)                 & (...)                &  10    \\
		J1518+4904                   & (...)                 & (...)             & $24.4289793809236(3)$ & $-0.0162263(12)$     &  11   \\
		J1930--1852                   & $185.52016047926(8)$ & $18.001(6)$      & (...)                 & (...)                &  12
	\end{tabular}
\tablefoot{
We follow the literature in giving either $P$ and $\dot{P}$ or $\nu$ and $\dot{\nu}$ for each pulsar, even though these are essentially the same quantities. The values in parentheses show the $1\sigma$ uncertainty in the preceding digits.\\
\tablefoottext{a}{Two components of the double-pulsar binary J0737--3039A/B.}
}
\tablebib{
    (1) \citet{Kramer_2021}; (2) \citet{Kramer_2006}; (3) \citet{Weisberg_2016}; (4) \cite{Ferdman_2020}; (5) \cite{Lynch_2018}; (6) \citet{Ferdman_2014}; (7) \citet{Fonseca_2014}; (8) \cite{Haniewicz_2021}; (9) \citet{Martinez_2017}; (10) \citet{Martinez_2015}; (11) \cite{Janssen_2008};  (12) \cite{Swiggum_2015}. 
	}
\end{table*}
\begin{table*}
    \centering
    \caption{Kinematic properties of the Galactic BNSs: the right ascension (R.A.) and declination (Dec.), the proper motion in right ascension ($\mu_{\alpha}$) and declination ($\mu_{\delta}$), the distance from the Sun ($D$), and the references (Ref.) for the listed values.}
    \label{tab2}
    \begin{tabular}{lcccccc}
    \hline\hline\\[-10pt]
    BNS pulsar & R.A. & Dec. & $\mu_{\alpha}$ & $\mu_{\delta}$ & $D$ & Ref. \\
    & (J2000) & (J2000) & [mas yr$^{-1}$] & [mas yr$^{-1}$] & [kpc] &  \\ 
		\hline\\[-10pt]
		J0737--3039A/B       & $07^{\rm h}37^{\rm m}51\fs 248115(10)$  & $-30\degr39'40\farcs70485(17)$  & $-2.57(3)$       & $2.08(4)$       & $0.74(6)$                 & 1\\ \\[-10pt]
		B1913+16            & $19^{\rm h}15^{\rm m}27\fs 99942(3)$  & $+16\degr06'27\farcs3868(5)$  & $-0.77(11)$    & $-0.01(14)$  & $4.1^{+2.0}_{-0.7}$ \hspace{3mm}\tablefootmark{a}               & 2\\ \\[-10pt]
		J1913+1102 & $19^{\rm h}13^{\rm m}29\fs 05365(9)$   & $+11\degr02'05\farcs7045(22)$    & $-3.0(5)$        & $-8.7(10)$      & $7.3(15)$\tablefootmark{b}       & 3\\ \\[-10pt]
		J0509+3801 & $05^{\rm h}09^{\rm m}31\fs 788(1)$\tablefootmark{c}     & $+38\degr01'18\farcs10(1)$\tablefootmark{c}   & $2.9(1)$ & $-5.9(3)$       & $4.2^{+1.6}_{-0.9}$ \hspace{3mm}\tablefootmark{a}              & 4, 5\\ \\[-10pt]
		J1756--2251   & $17^{\rm h}56^{\rm m}46\fs 633812(15)$   & $-22\degr51'59\farcs35(2)$   & $-2.42(8)$       & $0(20)$         & $0.73^{+0.38}_{-0.15}$ \hspace{3mm}\tablefootmark{a, d}             & 6, 7 \\ \\[-10pt]
		B1534+12            & $15^{\rm h}37^{\rm m}09\fs 961730(3)$ & $+11\degr55'55\farcs43387(6)$ & $1.482(7)$       & $-25.285(12)$   & $1.05(5)$\tablefootmark{c}                 & 8\\ \\[-10pt]
		J1829+2456 & $18^{\rm h}29^{\rm m}34\fs 66838(6)$  & $+24\degr56'18\farcs2007(12)$  & $-5.51(6)$       & $-7.82(8)$       & $1.1(3)$\tablefootmark{b}        & 9\\ \\[-10pt]
		J1411+2551          & $14^{\rm h}11^{\rm m}18\fs 866(3)$   & $+25\degr51'08\farcs39(7)$  & $-3(12)$         & $-4(9)$         & $1.1(2)$\tablefootmark{b}        & 10\\ \\[-10pt]
		J0453+1559          & $04^{\rm h}53^{\rm m}45\fs 41368(5)$   & $+15\degr59'21\farcs3063(59)$  & $-5.5(5)$        & $-6.0(42)$      & $0.6(4)$\tablefootmark{b}        & 11\\ \\[-10pt]
		J1518+4904          & $15^{\rm h}18^{\rm m}16\fs 79817(2)$\tablefootmark{c}  & $+49\degr04'34\farcs1132(2)$\tablefootmark{c} & $-0.69(3)$       & $-8.53(8)$      & $0.81(2)$                 & 12\\ \\[-10pt]
		J1930--1852 & $19^{\rm h}30^{\rm m}29\fs 7156(7)$  & $-18\degr51'46\farcs27(6)$  & $4.3(2)$         & $-5.2(4)$       & $4.6^{+2.4}_{-1.4}$ \hspace{3mm}\tablefootmark{a}   & 5, 13
	\end{tabular}
        \tablefoot{
        The values in parentheses are the $1\sigma$ uncertainties in the preceding digits. If the distribution is not symmetric, we show the $68\%$ uncertainty in the upper and lower half of the distribution \citep[cf.\ appendix A of][]{Disberg_2023}.\\
        \tablefoottext{a}{Non-Gaussian probability distributions \citep[similar to e.g.][]{Verbiest_2012}.}
        \tablefoottext{b}{Weighted means of the DM-based distances from the Galactic free-electron density models of \cite{Yao_2017} and \cite{Cordes_2002}, as computed by \cite{Ding_2024}.}
        \tablefoottext{c}{Conservatively adopted uncertainties.}
        \tablefoottext{d}{Determined using \href{http://psrpop.phys.wvu.edu/LKbias/}{http://psrpop.phys.wvu.edu/LKbias/}.}
        }
        \tablebib{
        (1) \citet{Kramer_2021}; (2) \citet{Deller_2018}; (3) \citet{Ferdman_2020}; (4) \citet{Lynch_2018}; (5) \citet{Ding_2024}; (6) \citet{Ferdman_2014}; (7) \citet{Verbiest_2012}; (8) \citet{Fonseca_2014}; (9) \citet{Haniewicz_2021}; (10) \citet{Martinez_2017}; (11) \citet{Martinez_2015}; (12) \citet{Ding_2023}; (13) \citet{Swiggum_2015}.
        }
\end{table*}
\begin{table*}
\centering
\caption{Estimated time-scales of the Galactic BNS pulsars: the characteristic age ($\tau_{\text{c}}$, Eq.\ \ref{eq2}), kinematic age ($\tau_{\text{kin}}$) for the first ($\chi_1$) and second ($\chi_2$) disc crossing, assuming either GC-isotropy or LSR-isotropy, and the merger time ($\tau_{\text{gw}}$).}
\label{tab3}
\begin{tabular}{lcccccc}
\hline\hline\\[-10pt]
BNS pulsar                    &  $\tau_{\text{c}}$                    & \multicolumn{4}{c}{$\tau_{\text{kin}}$}                                                                                                        & $\tau_{\text{gw}}$\tablefootmark{a}  \\
                              &   &   \multicolumn{2}{c}{$\chi_1$}               & \multicolumn{2}{c}{$\chi_2$}                 &                                      \\
 & & GC-isotropy & LSR-isotropy & GC-isotropy & LSR-isotropy & 
                              \\
                              & [Myr]               & [Myr]                             & [Myr]                           & [Myr]                                & [Myr]                             & [Myr]                                \\ \hline\\[-10pt]
J0737--3039A\tablefootmark{b} & $204.49247(7)$                           & \multirow{2}{*}{$66^{+41}_{-38}$} & \multirow{2}{*}{$40^{+4}_{-5}$} & \multirow{2}{*}{$503^{+285}_{-338}$} & \multirow{2}{*}{$96^{+14}_{-14}$} & \multirow{2}{*}{$85$}                \\
J0737--3039B\tablefootmark{b} & $49.3(4)$                                                       &                                   &                                 &                                      &                                   &                                      \\ \\[-10pt]
B1913+16                      & $108.596(4)$                             & $10^{+2}_{-2}$                    & $7^{+1}_{-1}$                   & $28^{+6}_{-2}$                       & $37^{+11}_{-7}$                   & $301$                                \\ \\[-10pt]
J1913+1102                    & $2457(1)$                                & $45^{+45}_{-15}$                  & $51^{+46}_{-16}$                & $68^{+69}_{-21}$                     & $105^{+109}_{-39}$                 & $465$                                \\ \\[-10pt]
J0509+3801                    & $153.02(4)$                              & $4^{+1}_{-1}$                     & $4^{+1}_{-1}$                   & $72^{+27}_{-19}$                     & $63^{+37}_{-15}$                  & $579$                                \\ \\[-10pt]
J1756--2251                   & $443.494(1)$                             & $3^{+31}_{-2}$                    & $2^{+34}_{-2}$                  & $58^{+63}_{-25}$                     & $59^{+27}_{-15}$                  & $1457$                               \\ \\[-10pt]
B1534+12                      & $248.0794(3)$                            & $64^{+19}_{-46}$                  & $55^{+13}_{-40}$             & $136^{+73}_{-45}$                    & $105^{+67}_{-17}$                 & $2735$                               \\ \\[-10pt]
J1829+2456                    & $13\,152(6)$                             & $32^{+1}_{-4}$                    & $13^{+4}_{-3}$                  & $51^{+35}_{-4}$                      & $62^{+18}_{-7}$                   & $55\,375$\tablefootmark{c}            \\ \\[-10pt]
J1411+2551                    & $10\,300(600)$                           & $42^{+84}_{-35}$                  & $32^{+28}_{-17}$                & $164^{+140}_{-79}$                   & $107^{+76}_{-32}$                 & $465\,446$\tablefootmark{c}           \\ \\[-10pt]
J0453+1559                    & $3900(2)$                                & $10^{+138}_{-8}$                 & $15^{+20}_{-7}$                 & $86^{+137}_{-32}$                    & $81^{+16}_{-15}$                  & $1\,456\,721$\tablefootmark{c}         \\ \\[-10pt]
J1518+4904                    & $23\,870(2)$\tablefootmark{c} & $69^{+27}_{-49}$                  & $24^{+13}_{-9}$                 & $158^{+67}_{-66}$                    & $85^{+11}_{-2}$                    & $8\,826\,539$\tablefootmark{c}         \\ \\[-10pt]
J1930--1852                   & $163.40(6)$                              & $11^{+3}_{-2}$                    & $9^{+2}_{-1}$                   & $49^{+89}_{-14}$                     & $66^{+77}_{-23}$                  & $10^{8}$ \hspace{3mm}\tablefootmark{c}
\end{tabular}
\tablefoot{
Quantities are determined through a Monte Carlo estimation, as described in Sect.\ \ref{sec2}. The median values of the resulting distributions and their uncertainties are given similar to Table \ref{tab2}.\\
\tablefoottext{a}{Lower bounds of the values determined following \citet{Peters_1964}, similarly to \citet{Gaspari_2024}.}
\tablefoottext{b}{Two components of the double-pulsar binary J0737-3039A/B, which share the same $\tau_{\text{kin}}$ and $\tau_{\text{gw}}$.}
\tablefoottext{c}{Exceeds a Hubble time.}
}
\end{table*}
\subsection{Spin-down ages}
\label{sec2.1}
The first quantity related to the ages of the BNS systems uses the spin properties of the observed pulsars in the binaries. During their formation pulsars get spun-up to high frequencies, corresponding to an initial period $P_0$. The spin-down time it takes for the pulsar to slow down to a period $P$, given a certain braking-index $n$, is then given by
\begin{equation}
    \label{eq1}
    \tau_{\text{sd}}=\dfrac{P}{(n-1)\dot{P}}\left[1-\left(\dfrac{P_0}{P}\right)^{n-1}\right]\quad.
\end{equation}
The characteristic age of the pulsar ($\tau_{\text{c}}$) is determined by assuming the braking is caused by pure dipole radiation, meaning $n=3$ -- while multipole radiation results in values of $n>3$, which could reach values op to $n=5$ if the braking is dominated by gravitational radiation \citep[e.g.][]{Camilo_1994,Ferrario_2007} -- and the initial period is small compared to the current period. This gives, as a function of $P$ and $\dot{P}$ or as a function of the frequency $\nu=P^{-1}$ and its derivative $\dot{\nu}=-\dot{P}P^{-2}$ \citep[e.g.][]{Shapiro_1983,Arzoumanian_1999}:
\begin{equation}
    \label{eq2}
    \tau_{\text{c}}=\dfrac{P}{2\dot{P}}=-\dfrac{\nu}{2\dot{\nu}}\quad.
\end{equation}
For some pulsars, $\tau_{\text{c}}$ seems to approximate their true age, for example because it aligns with an \textquotedblleft expansion age\textquotedblright\  for a supernova remnant based on ejecta velocities \citep[e.g.][]{Wyckoff_1997}, though it is not always a good estimate of the true age of a pulsar \citep{Jiang_2013,Zhang_2022}. For a millisecond pulsar (MSP) in a BNS, for instance, the assumptions that $P_0 \ll P$ and $n=3$ may be inaccurate. These MSPs are thought to have been spun-up by accreting mass from their companion, through which they were \textquotedblleft recycled\textquotedblright\ to millisecond periods \citep[e.g.][]{Bhattacharya_1991}.\\
\begin{figure*}
    \centering
    \includegraphics[width=18cm]{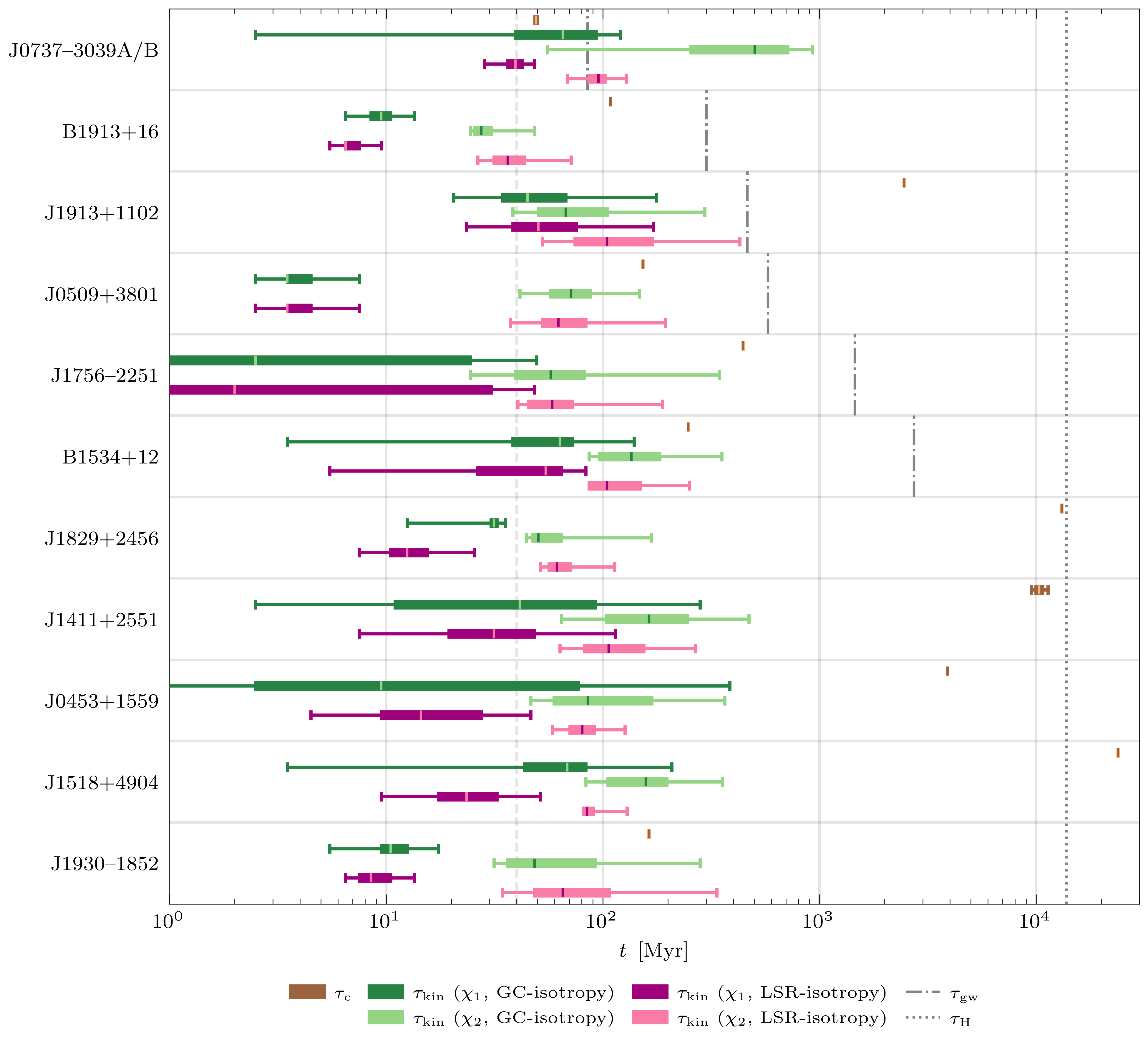}
    \caption{Age-related quantities for the Galactic BNSs as discussed in Sect.\ \ref{sec2}: the characteristic age $\tau_{\text{c}}$ (brown, through Eq.\ \ref{eq2}) and the kinematic ages $\tau_{\text{kin}}$ assumming GC-isotropy (dark green for $\chi_1$ and light green for $\chi_2$), and kinematic ages assuming LSR-isotropy (dark purple for $\chi_1$ and light purple for $\chi_2$). The boxes extend from the first to third quartile ($25\%$ to $75\%$ percentile) of the distributions, while the whiskers show the $5\%$ and $95\%$ percentiles and the different coloured lines show the medians. The figure also contains lines corresponding to $\tau_{\text{gw}}$ (dash-dotted lines), the Hubble time $\tau_{\text{H}}=13.8$ Gyr \citep[dotted line,][]{Komatsu_2011}, as well as a background line at $t=40$ Myr (dashed line, relevant for Sect.\ \ref{sec3}). The median values of these distributions and their uncertainties are listed in Table \ref{tab3}. For J0737--3039A/B we show $\tau_{\text{c}}$ of J0737--3039B.}
    \label{fig1}
\end{figure*}
\indent In order to estimate $\tau_{\text{c}}$, we constructed Gaussian distributions corresponding to the observed values of $P$ and $\dot{P}$ or $\nu$ and $\dot{\nu}$ and their uncertainties (as given in Table \ref{tab1}). Then, we sampled these distributions $10^3$ times for each BNS and for each instance determined $\tau_{\text{c}}$ through Eq.\ \ref{eq2}. The resulting distributions of the characteristic ages are shown in Fig.\ \ref{fig1}, and their median values and uncertainties are given in Table \ref{tab3}. The distributions in Fig.\ \ref{fig1} show that $\tau_{\text{c}}$ is well constrained, due to the precision of $P$ and $\dot{P}$ measurements, and does not differ from values that can be found in literature \citep[e.g.][]{Kargaltsev_2006,Andrews_2019}. It remains, however, difficult to determine the true ages of these pulsars based on a spin-down age, in particular due to the (binary) evolutionary history of the MSPs. This is, for instance, reflected by the fact that the median $\tau_{\text{c}}$ of J1518+4904 exceeds a Hubble time. Nevertheless, \citet{Maoz_2024} argue that $\tau_{\text{c}}$ is a \textquotedblleft reliable indicator of age,\textquotedblright\ and that corrections for binary evolution are expected to be relatively small \citep[based on][]{Kiziltan_2010}.
\subsection{Kinematic ages}
\label{sec2.2}
As an additional way to investigate the ages of the Galactic BNSs, we estimated their kinematic ages. That is, a BNS should be formed through supernovae in the thin disc, at $z\approx0$ kpc, after which it is kicked into an orbit containing heights $|z|\ >0$. We can use the motion of the binaries away from or towards the disc to determine the last time they crossed $z=0$. This could potentially give a lower bound for their true age, since the binaries may have crossed the disc multiple times since their birth. However, even if a BNS has not crossed the disc multiple times in its kinematic history, there are many reasons why its kinematic age might differ (significantly) from its actual age. After all, a BNS receives two systemic kicks, after the SNe of each component, which means that it could already start to migrate through the Galaxy before receiving its final systemic kick. In addition, the kinematic ages are made uncertain by the unknown radial velocity with respect to the Solar System ($v_r$) and the unknown birth location of the binaries (which could be at small but non-zero $|z|$). For these reasons we do not consider the kinematic ages of the BNSs as accurate estimates of their actual ages, but since there are several studies concerning the kinematic ages of individual BNSs, we determined them here to give a complete overview.\\
\indent We estimated the galactocentric speeds for the Galactic BNSs in our sample. For these binaries, the sky locations, proper notions, and distances are given in Table \ref{tab2}. In order to determine the magnitude of their galactocentric velocity vector, we estimated the unknown velocity in the radial direction ($v_r$) through two different isotropy assumptions \citep[following the method of][]{Gaspari_2024}. Firstly, we assumed that the full galactocentric velocity is oriented isotropically. We refer to this kind of isotropy as \textquotedblleft GC-isotropy.\textquotedblright\ For a BNS where the proper motion and distance translates into a transverse velocity $v_t$, the GC-isotropic estimate of $v_r$ equals:
\begin{equation}
    \label{eq3}
    v_r = v_t\cot\theta\quad,
\end{equation}
where $\theta=\arccos u$ and $u$ is uniformly sampled between $0$ and $1$. We sampled $u$ $10^3$ times, with which we created a distribution of $v_r$ for each BNS. Secondly, we assumed that the peculiar velocity of the BNSs is oriented isotropically in the local standard of rest (LSR), which we refer to as \textquotedblleft LSR-isotropy.\textquotedblright\ If a BNS is located at a position where the velocity vector of the LSR equals $\vec{v}_{\text{LSR}}$, then the unknown $v_r$ is estimated by subtracting the 2D transverse part of the LSR velocity vector ($\vec{v}_{\text{LSR},\,t}$) from the BNS' $\vec{v}_t$ and adding the radial part of the LSR motion.
\begin{equation}
    \label{eq4}
    v_r=\left|\left|\vec{v}_t-\vec{v}_{\text{LSR},\,t}\right|\right|\hspace{.3mm}\cot\theta+v_{\text{LSR},\,r}\quad,
\end{equation}
where $\theta$ is defined similar to Eq.\ \ref{eq3}, meaning this also resulted in a distribution of $v_r$ containing $10^3$ values. The LSR-isotropy assumption is appropriate if the BNS receives low kicks, and its velocity is therefore dominated by the initial circular velocity of the LSR (as determined by the Galactic potential).\\
\indent We obtained a $v_r$ distribution from assuming either GC-isotropy or LSR-isotropy, and then used the BNS sky-locations, distances, and 2D proper motions of the BNS systems (which are given in Table \ref{tab2}) to construct Gaussian distributions based on their mean and uncertainties, for each binary. We sampled these distributions $10^3$ times and determined the 3D velocity vectors for each instance. Having obtained $10^3$ sets of positions and velocities for each BNS, we used the \lstinline{GALPY}\footnote{\href{http://github.com/jobovy/galpy}{http://github.com/jobovy/galpy}} \lstinline{v.1.9.0} package \citep{Bovy_2015} to integrate the orbits backwards in the Milky Way (MW) potential of \citet{McMillan_2017}, assuming that the circular velocity at the position of the Sun equals its azimuthal velocity \citep[i.e.\ $v_{\text{LSR}}(8.122\text{ kpc})=245.6$ km/s,][]{Gravity_2018}. We evaluated these orbits every Myr, up to $1$ Gyr, in order to determine the points at which they cross $z=0$. If $z(t_i)\cdot z(t_{i+1})<0$, meaning one has a positive $z$ and the other a negative one, we set the time of crossing as the average between these consecutive timestamps. Using this method, we determined the distributions of $\tau_{\text{kin}}$ for the first ($\chi_1$) and second ($\chi_2$) time each BNS crosses the disc (when tracing back their kinematic history), where we neglect orbits that do not cross within $1$ Gyr, or do so at galactocentric radii $R>30$ kpc.\\
\indent In Fig.\ \ref{fig1} we show the distributions of $\tau_{\text{kin}}$ for $\chi_1$ and $\chi_2$, assuming either GC-isotropy or LSR-isotropy. For most BNSs, except perhaps J0737--3039A/B and J1518+4904, the distributions of $\tau_{\text{kin}}$ for GC-isotropy and LSR-isotropy agree relatively well. For J0737--3039A/B, the characteristic age overlaps with $\tau_{\text{kin}}$ for $\chi_1$ assuming GC-isotropy, and $\chi_2$ assuming LSR-isotropy. In general, the characteristic ages of the binaries exceed their kinematic ages signficantly. While it is difficult to estimate the reliability of the spin-down ages, the difference of one or more orders of magnitude between the spin-down and the kinematic ages does imply they were likely formed at $\chi_{\geq3}$. In Table \ref{tab3} we list the median kinematic ages and their uncertainties. Also, the boxplots in Fig.\ \ref{fig1} hide structures such as potential bimodality in the $\tau_{\text{kin}}$ distributions, which is why we show the disc-crossings in Appendix \ref{app.A}. For most binaries, and particularly for LSR-isotropy, the $\tau_{\text{kin}}$ estimates describe the disc crossings relatively well.\\
\indent \citet{Willems_2004} discussed the kinematic age of J0737--3039A/B, depending on assumed values of $v_r$ and the angle of the 1D proper motion on the sky ($\Omega$). They conclude that for almost all values of $v_r$ and $\Omega$, $\tau_{\text{kin}}$ for $\chi_1$ is less than $100$ Myr and $\tau_{\text{kin}}$ for $\chi_2$ exceeds $20$ Myr, both of which match our findings. In addition, \citet{Lorimer_2005} employed spin-down models to constrain the age of J0747--3039A/B to $30-70$ Myr, which also coincides with the region of overlap between $\tau_{\text{c}}$ and $\tau_{\text{kin}}$ we find. \citet{Willems_2006} found that certain values of $v_r$ give rise to $\tau_{\text{kin}}\sim5$ Myr for $\chi_1$. These solutions can also be seen in Appendix \ref{app.A}. For B1534+12, \citet{Willems_2004} found that for $\chi_1$, $\tau_{\text{kin}}$ spans a range between $1$ and $210$ Myr, and for $\chi_2$ the kinematic age is at least $90$ Myr, both of which are compatible with our $\tau_{\text{kin}}$ distributions. \citet{Arzoumanian_1999}, in turn, found that B1534+12 has likely crossed $z=0$ more than once, which is supported by the fact that $\tau_{\text{c}}$ likely exceeds $\tau_{\text{kin}}$ for $\chi_2$. \citet{Arzoumanian_1999} also noted that the BNS B1913+16 can be constrained well kinematically, since it has a low altitude ($z$) and is moving away from the disc. They estimated $\tau_{\text{kin}}$ for $\chi_1$ to be $\sim5$ Myr, which is only slightly below our estimates, and $\tau_{\text{kin}}$ for $\chi_2$ to be $60-80$ Myr, which exceeds our estimates for $\chi_2$ but aligns better with the characteristic age. The estimates of \citet{Willems_2004} are similar, with $\tau_{\text{kin}}\simeq2-4$ Myr for $\chi_1$ and $\tau_{\text{kin}}\gtrsim55$ Myr for $\chi_2$ \citep[and are also in agreement with the analysis of][]{Wex_2000}.\\
\indent Moreover, we considered the merger times ($\tau_{\text{gw}}$) of the binaries. These are of course not age-estimates themselves, since they give the amount of time needed for these binaries to merge in the future. However, the rate at which the orbital separation of the binaries shrink -- due to the emission of gravitational waves -- strongly depends on the current orbital separation, meaning the binaries spend most of their time close to their initial orbital separation. This means that, on a population level (but not necessarily on an individual level), we expect binaries with longer merger times to be older than systems with younger merger times \citep[for further discussion see][]{Maoz_2024}. We show the $\tau_{\text{gw}}$ values in Fig.\ \ref{fig1} and list them in Table \ref{tab3} \citep[which agree with values given by e.g.][]{Faulkner_2005,Tauris_2017}. The merger times of several BNSs exceed a Hubble time and are significantly longer than the merger times of the other binaries. With the exception of J1930--1852, these binaries indeed have the largest values of $\tau_{\text{c}}$.
\section{Speeds}
\label{sec3}
\begin{figure}
    \resizebox{\hsize}{!}{\includegraphics{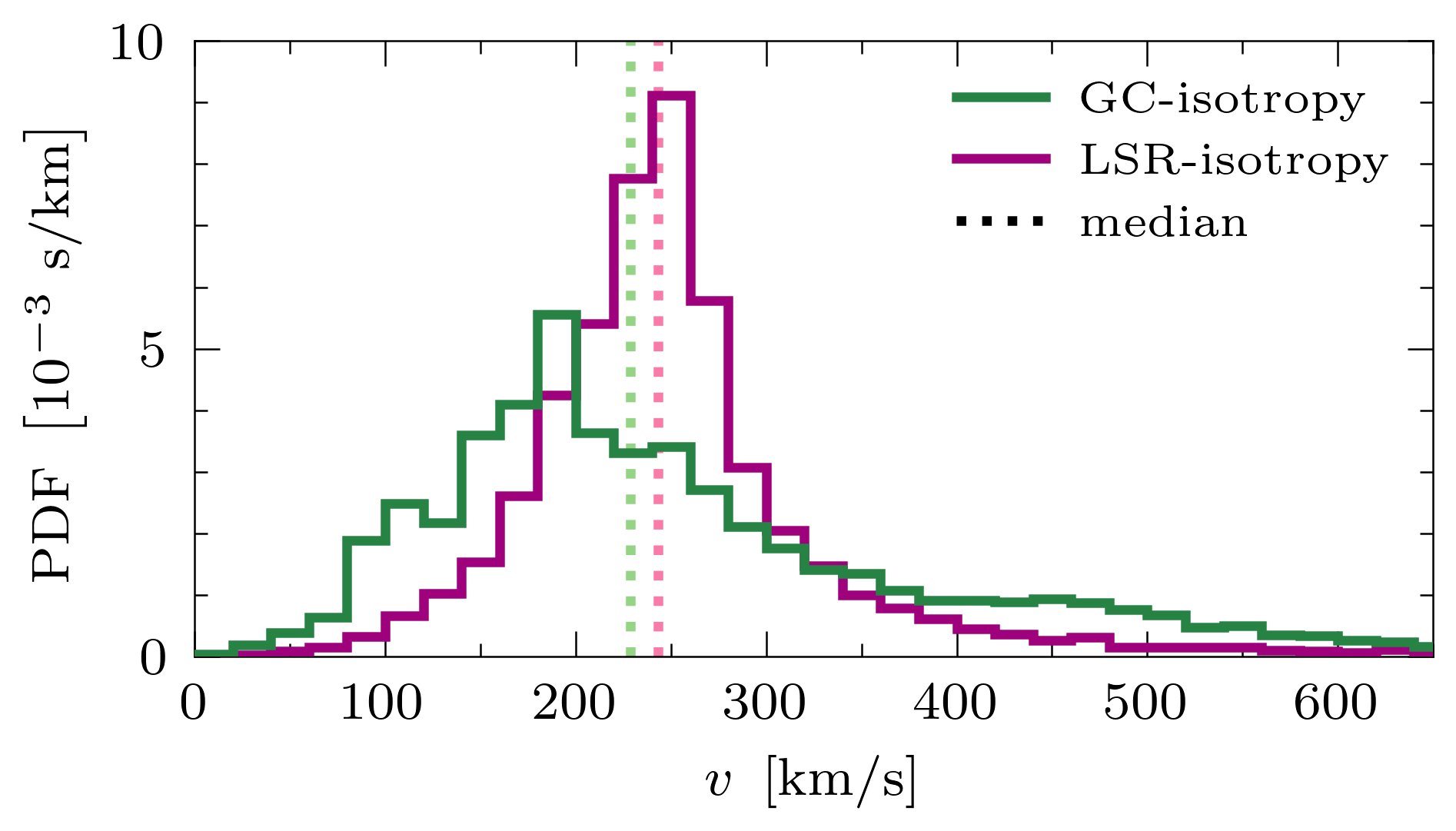}}
    \caption{Total distributions of the magnitudes of the Monte Carlo estimated present-day galactocentric velocity vectors of the BNSs ($v$), shown in normalised histograms with bins of $20$ km/s, for GC-isotropy (green) and LSR-isotropy (purple), together with dotted lines showing the corresponding median speeds.}
    \label{fig2}
\end{figure}

We are interested in the magnitude of kicks the Galactic BNSs experienced at their formation. As a first attempt to determine the kick velocities of the binaries, we investigated their current speeds, as estimated through the method discussed in Sect.\ \ref{sec2.2}. For each binary, we obtained $10^3$ positions and velocities, resulting in a distribution of the present-day galactocentric speeds for each binary. In Fig.\ \ref{fig2}, we show the combined speed distributions of the BNS systems in our sample \citep[for the velocity distributions for some of the individual BNSs, see][]{Gaspari_2024}. For both GC-isotropy and LSR-isotropy, the speed distributions peak around $200-250$ km/s. Since this is comparable to the circular velocity of the Solar System, one could argue that this means the binaries are likely to have experienced low kicks. However, \citet{Disberg_2024} show that this is not necessarily the case. Here, we will briefly recreate and expand their argument for why it is difficult to infer kicks based on current (galactocentric) speeds. In order to do this, we simulated kicked objects in the Galactic potential (Sect.\ \ref{sec3.1}) and compared the results to the Galactic BNSs (Sect.\ \ref{sec3.2}).
\subsection{Simulation}
\label{sec3.1}
\begin{figure*}
    \sidecaption
    \includegraphics[width=12cm]{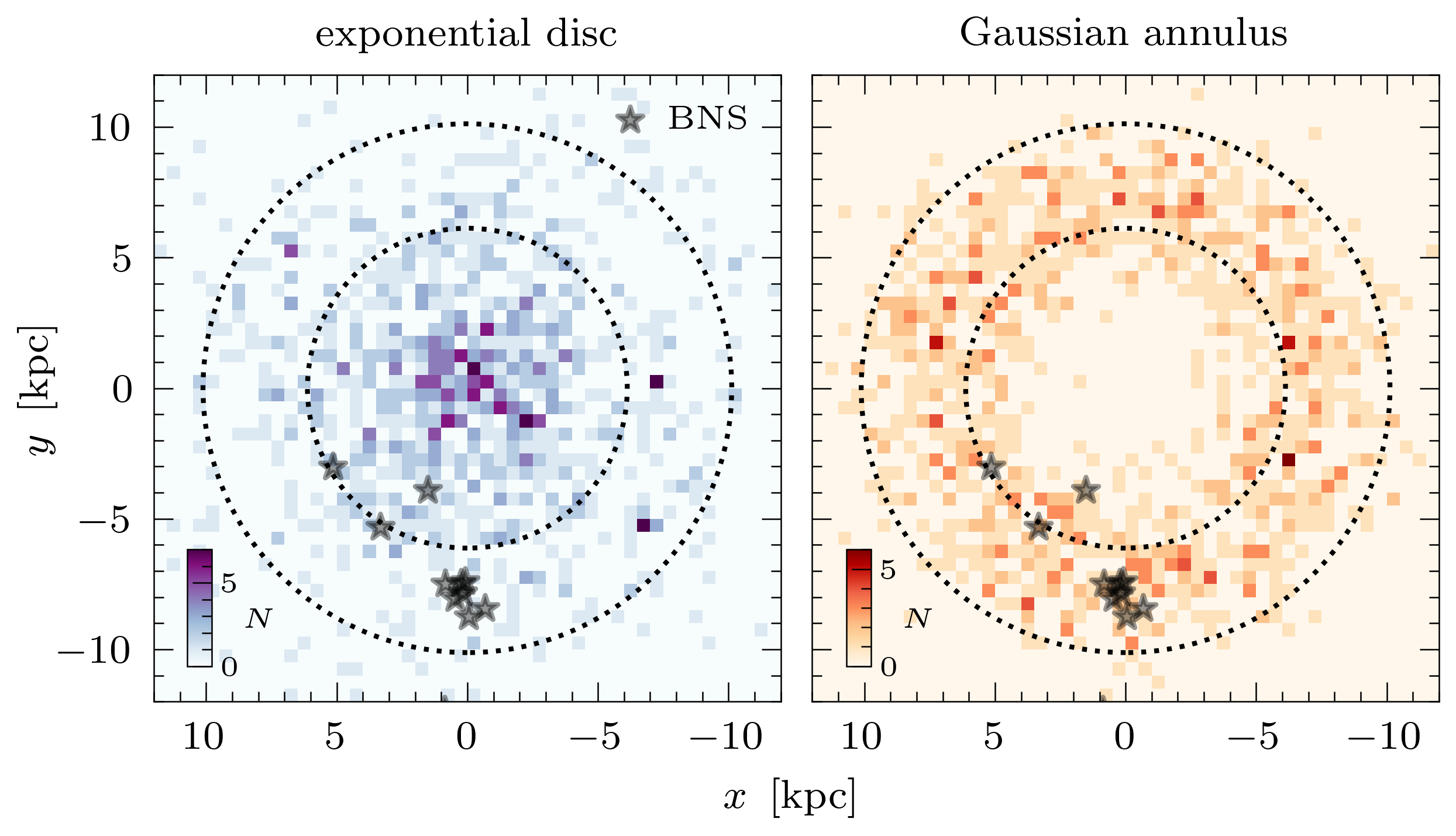}
    \caption{Priors for the initial positions in our simulation, shown by seeding $10^3$ points and displaying their density in normalised 2D histograms with bins of $0.5$ kpc. The left panel shows the exponential disc prior from \citet{Disberg_2024}, and the right panel shows the Gaussian annulus prior as defined in Eq. \ref{eq5}. The black stars show the positions of the Galactic BNSs, determined as the mean of the Monte Carlo distributions described in Sect.\ \ref{sec2.2} \citep[cf.\ ][]{Gaspari_2024}. The dotted lines show $R=R_{\sun}\pm2$ kpc, which trace the edges of the solar neighbourhood \citep[as defined by][]{Disberg_2024}.}   
    \label{fig3}
\end{figure*}

The argument of \citet{Disberg_2024} is based on a Monte Carlo simulation of objects moving through the Galactic potential after receiving a kick. First, they seeded point-masses in an assumed prior distribution which is described by an exponential disc \citep[convolved with a prescription for the spiral arms, using the work of][]{Chrimes_2021}. Here, we created a similar simulation, using the prior distribution from \citet{Disberg_2024}, and compared it with a simulation using a different prior. That is, we compared it with a simulation using a Gaussian annulus as prior distribution for the positions of the objects. Such a distribution was first proposed by \citet{Faucher_2006}, and adopted by \citet{Sartore_2010} to fit the pulsar distribution found by \citet{Yusifov_2004}. They adopted the following Gaussian annulus distribution:
\begin{equation}
    \label{eq5}
    \rho(R)\propto\exp\left(-\dfrac{\left(R-7.04\text{ kpc}\right)^2}{2\left(1.83\text{ kpc}\right)^2}\right)\quad,
\end{equation}
where $R$ is the galactocentric radius. In Fig.\ \ref{fig3}, we seeded $10^3$ points in both the exponential disc prior and the Gaussian annulus prior, and show their positions (together with the positions of the BNSs). The figure shows the significant difference between these priors: in the exponential disc most points are located near the Galactic centre, while in the Gaussian annulus most points are located at $R\approx R_{\sun}$.\\
\indent Similarly to Fig.\ \ref{fig3}, we seeded $10^3$ point-masses in each prior (determining their initial positions), gave them an initial velocity corresponding to the circular velocity at their initial $R_{0}$ -- as dictated by the MW potential of \citet{McMillan_2017} -- and added a kick velocity which is isotropically oriented \citep[for a more detailed description, see][]{Disberg_2024}. We sampled the magnitude of the kick velocity from a Maxwellian distribution, and repeated this simulation for several Maxwellians, choosing values for $\sigma$ of $10$, $20$, $50$, $100$, $200$, and $500$ km/s \citep[cf.][who estimated the $\sigma$ for isolated pulsars to be $265$ km/s]{Hobbs_2005}. Having established the initial position and initial velocity vector of the $10^3$ point masses, we computed their trajectories using \lstinline{GALPY v.1.9.0} \citep{Bovy_2015} and the MW potential of \citet{McMillan_2017}, evaluating the positions and speeds of the point-masses every Myr for $200$ Myr. Since we want to compare the speeds of the kicked objects to the observed speeds of the BNSs, we selected the objects that are, at a certain time, in the solar neighbourhood. We adopted the solar neighbourhood definition of \citet{Disberg_2024}, who -- motivated by the cylindrical symmetry of the system -- initialised the orbit of the Sun $12$ times, each one azimuthally rotated by $2\pi/12$, and evolved these together with the point-masses. Then, at each point in time, they evaluated the speeds of objects that are within $2$ kpc of one of the solar obits.
\begin{figure*}
    \centering
    \includegraphics[width=18cm]{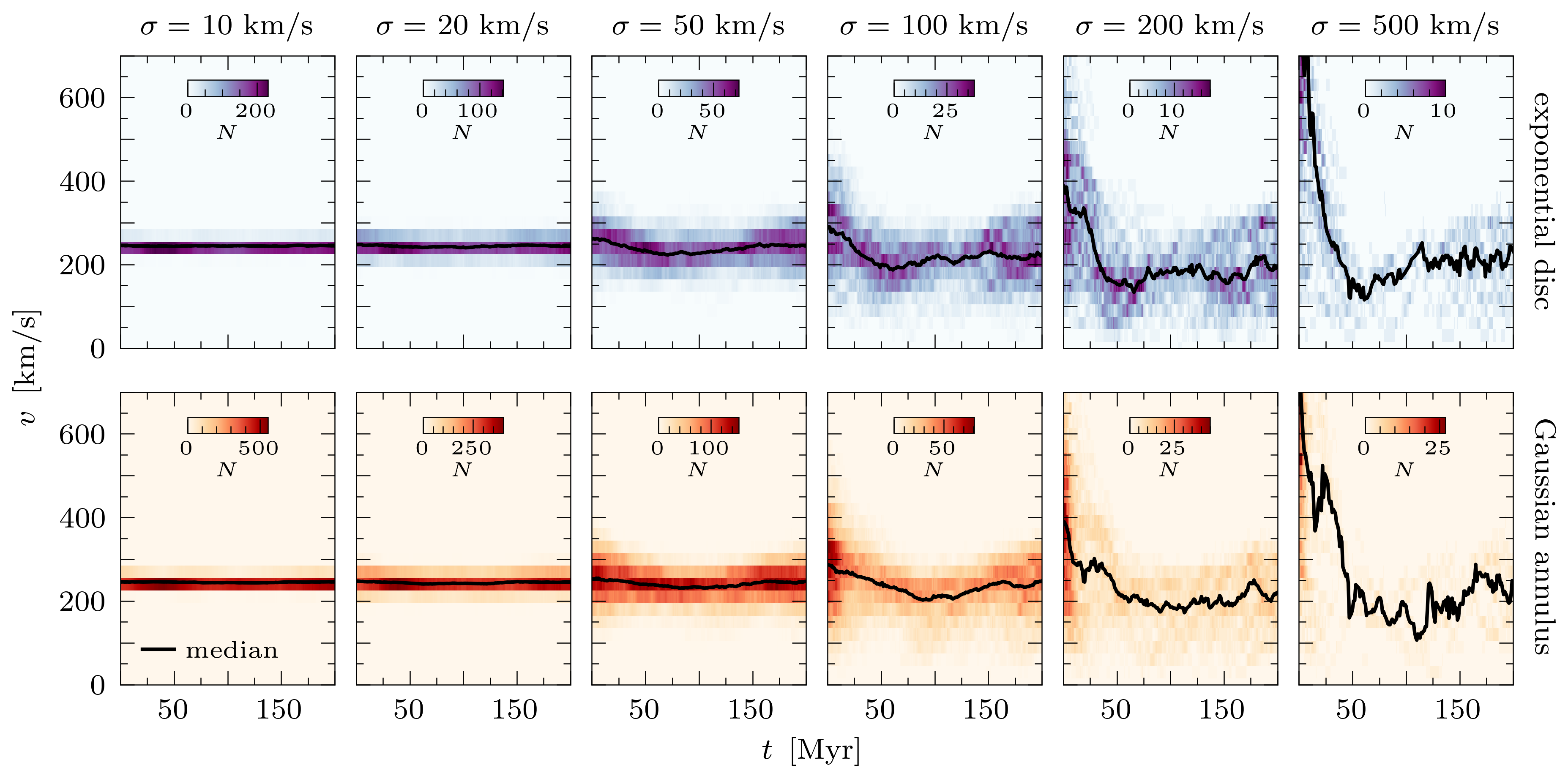}
    \caption{Evolution of the galactocentric speeds ($v$) of the objects in the solar neighbourhood, for the different Maxwellian kick distributions ($\sigma=10,\ 20,\ 50,\ 100,\ 200,\ \text{and }500$ km/s, for each column), shown in 2D histograms with velocity bins of $30$ km/s and time bins of $1$ Myr. The top row shows the results using the exponential disc prior \citep[cf.\ Appendix B of][]{Disberg_2024}, while the bottom row shows the results using the Gaussian annulus prior. The black lines correspond to the median speed at each time bin. We note that the colour-scale differs for each panel, because the number of objects in the solar neighbourhood differs for each kick distribution.}
    \label{fig4}
\end{figure*}
\subsection{Deceleration}
\label{sec3.2}
We repeated the simulation described above for each Maxwellian ($\sigma=10$, $20$, $50$, $100$, $200$, and $500$ km/s), and obtained the galactocentric speeds shown in Fig.\ \ref{fig4}. In accordance with the results of \citet{Disberg_2024}, we find that after a certain amount of time the objects in the solar neighbourhood have decreased speeds: they have been decelerated by the Galactic potential. That is, for higher kicks the objects that we see in the solar neighbourhood have more eccentric orbits, while we are more likely to observe them closer to their apocentre where they have a lower speed \citep[corresponding to the asymmetric drift found by][]{Hansen_1997}. Because of this, the speeds of older objects in the solar neighbourhood are reduced, and have a median value of $\sim150-250$ km/s, independent of the kick distribution. The speed distributions of these older objects do differ in their spread, paradoxically resulting in the lowest speeds being observed for the highest kicks. For young objects, the speeds do depend on the kicks, but these can include objects that become unbound, and are therefore not seen in the solar neighbourhood at later times. Moreover, we find no significant difference between the exponential disc and the Gaussian annulus prior, when it comes to these speeds. After all, orbits with identical eccentricity can be formed at different values of $R_0$ (albeit with a different kick magnitude and orientation).\\
\indent We do find that in general there tend to be more objects in the solar neighbourhood for the Gaussian annulus prior, because in this prior more objects are formed near or in the solar neighbourhood. However, the relative amount of objects that cross the solar neighbourhood in our simulation may not be representative of the observable BNSs in the Galaxy, because these are only observable during the limited time they are visible as a radio pulsar. Also, after a certain amount of time the BNS merges and is no longer observable as a BNS, and the merger times may -- to some degree -- depend on the kick magnitude. Nevertheless, while these effects may be important in estimating the formation rates of BNSs from the observed Milky Way population, it has little impact on our work, which seeks to infer the kicks.\\
\begin{figure*}
    \sidecaption
    \includegraphics[width=12cm]{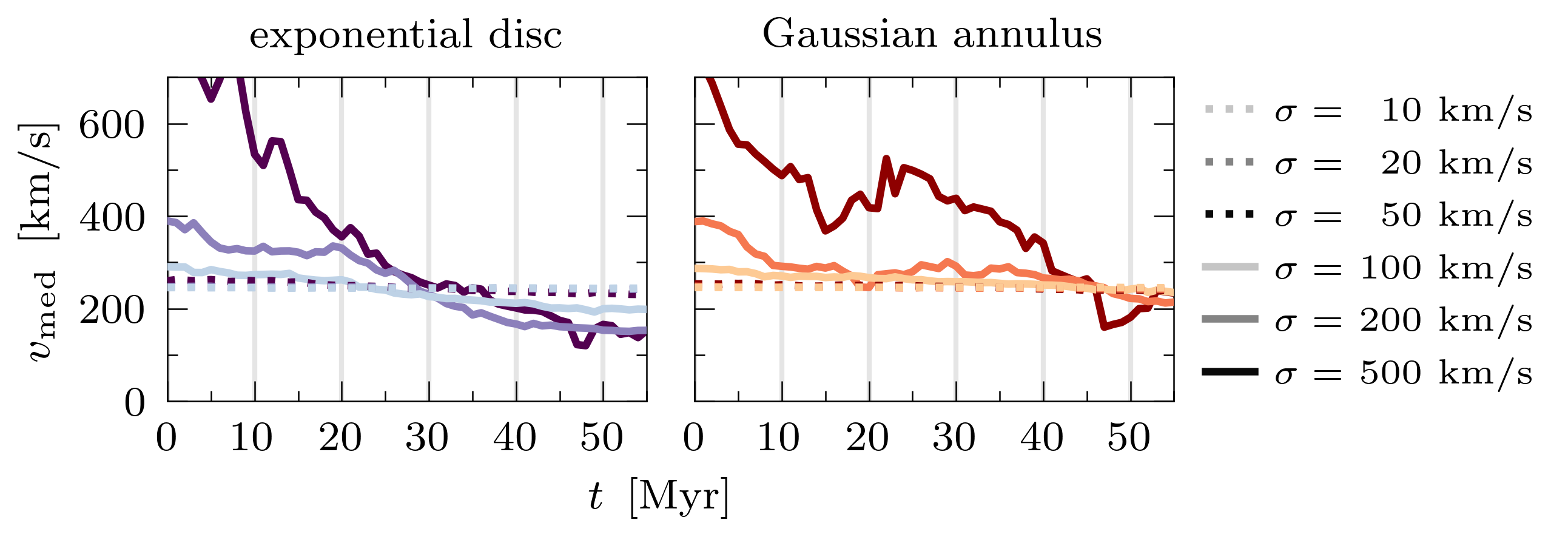}
    \caption{Medians of the galactocentric speeds ($v_{\text{med}}$) as shown in Fig.\ \ref{fig4}, for kick distributions described by $\sigma=10$ km/s (light dotted), $20$ km/s, (dotted), $50$ km/s (dark dotted), $100$ km/s (light), $200$ km/s, and $500$ km/s (dark).}   
    \label{fig5}
\end{figure*}
\begin{figure*}
    \centering
    \includegraphics[width=18cm]{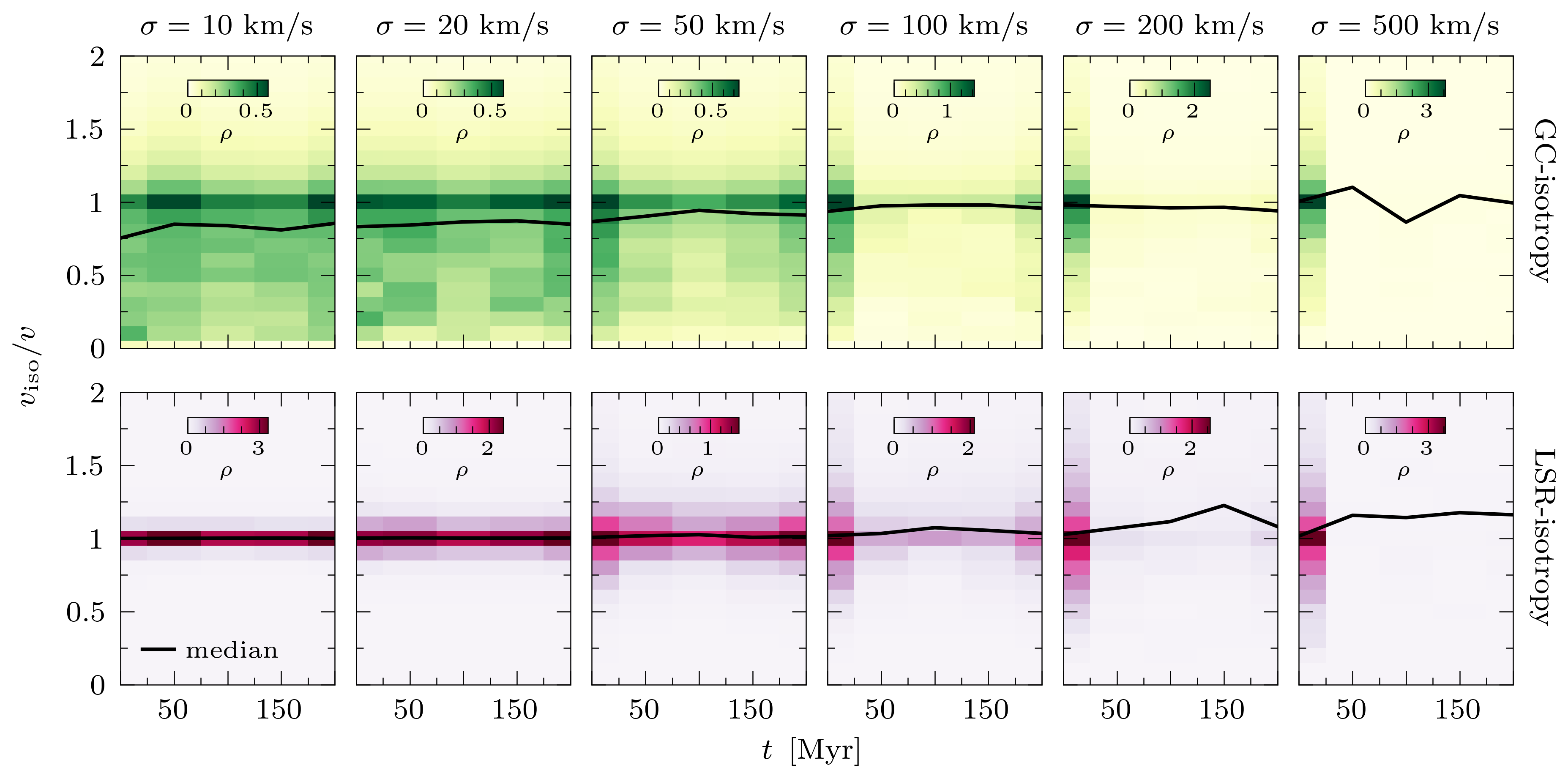}
    \caption{Accuracy of the velocity distributions determined through isotropy assumptions ($v_{\text{iso}}$) for the objects in our simulation (of which the speeds are shown in Fig.\ \ref{fig4}) that are in the solar neighbourhood at a certain point in time. This accuracy is estimated by dividing these distributions by the actual speeds ($v$) and evaluating $v_{\text{iso}}/v$ every $50$ Myr, for GC-isotropy (top row, through Eq.\ \ref{eq3}) and LSR-isotropy (bottom row, through Eq.\ \ref{eq4}), and for Maxwellian kick distributions with different values of $\sigma$ (columns), shown in normalised 2D histogram with $v_{\text{iso}}/v$ bins of $0.1$ and time bins of $50$ Myr, in units of $\left(500\text{ Myr}\right)^{-1}$. The black lines show the medians of the distributions.}
    \label{fig6}
\end{figure*}
\begin{figure*}
    \centering
    \includegraphics[width=18cm]{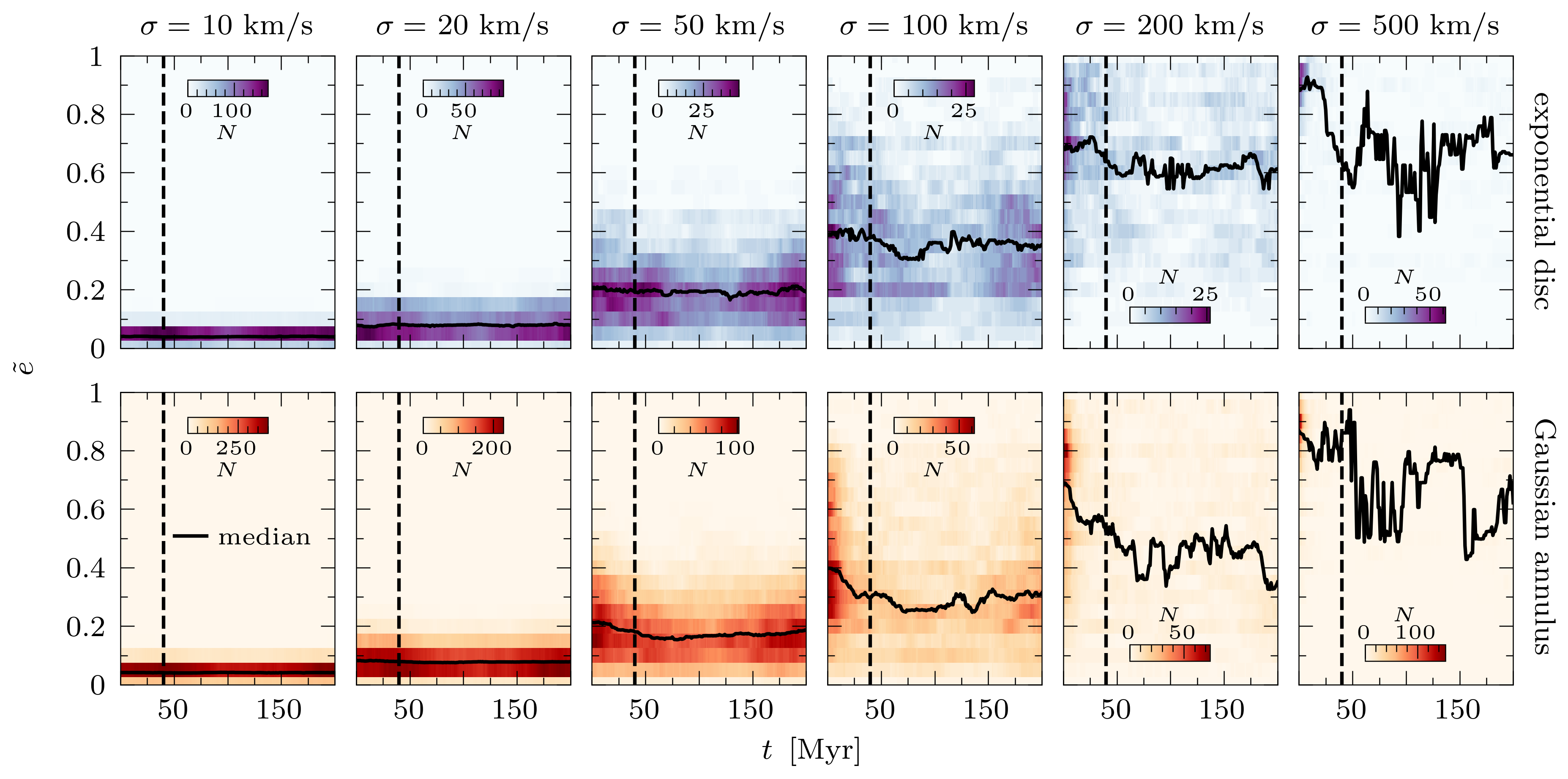}
    \caption{Evolution of the eccentricities ($\tilde{e}$, Eq.\ \ref{eq6}) of the objects in the solar neighbourhood, for the different Maxwellian kick distributions ($\sigma=10,\ 20,\ 50,\ 100,\ 200,\ \text{and }500$ km/s, for each column), shown in 2D histograms with eccentricity bins of $0.05$ and time bins of $1$ Myr. The top row shows the results using the exponential disc prior, while the bottom row shows the results using the Gaussian annulus prior. The black dashed lines show $t=40$ Myr, approximating the timescale after which the median velocities have been decelerated (as shown in Fig.\ \ref{fig5}). Similarly to Fig.\ \ref{fig4}, we note that the colour-scale differs for each panel, because the number of objects in the solar neighbourhood differs for each kick distribution.}
    \label{fig7}
\end{figure*}
\indent Figure\ \ref{fig4} shows how the median speeds decrease rapidly for the different kick distributions, obtaining similar values independent of their initial kicks. In Fig.\ \ref{fig5}, we plot these medians together, for the exponential disc and the Gaussian annulus prior, to show the timescales of this effect. The figure shows that for the exponential disc prior, the medians for all $\sigma$ obtain similar values sometime between $20$ and $30$ Myr \citep[as also found by][]{Disberg_2024}, and that for the Gaussian annulus prior the medians become similar around $40$ Myr. In this work, we conservatively adopt $40$ Myr as the timescale above which we expect the objects in the solar neighbourhood to be decelerated and have similar median speeds ($\sim150-250$ km/s), and therefore provide little information about the kick distribution.\\
\indent In Fig.\ \ref{fig1}, we show a line at $40$ Myr. We argue that for the Galactic BNSs, it is reasonable to estimate them \emph{all} to be older than $\sim40$ Myr, since Fig.\ \ref{fig1} and Table \ref{tab3} show that (1) the characteristic ages of the BNSs are well above $40$ Myr, meaning that their true ages retain values above this limit even if $\tau_{\text{c}}$ is a considerable overestimation (with the only exception being $\tau_{\text{c}}$ for J0737--3039B, but this is not a MSP), and (2) the kinematic ages of the binaries indicate that there is only one likely disc-crossings below $40$ Myr (exception being B1913+16, which could have two). Because the Galactic BNSs are all likely to be older than $40$ Myr, we expect them to have median speeds around $\sim150-250$ km/s, regardless of the kick distribution. This corresponds to the galactocentric speeds we find, as shown in Fig.\ \ref{fig2}, with median speeds $\sim200-250$ km/s. If we could estimate $v_r$ more precisely and get more accurate estimates for the BNSs velocities, we could perhaps differentiate between the speed distributions shown in Fig.\ \ref{fig4}, but currently the uncertainty in the BNS velocity estimates prevents us from doing so. We therefore conclude that, based on the galactocentric speeds of the BNSs alone, we are not able to constrain their kicks.\\
\indent Moreover, the simulation of which we show the speeds in Fig.\ \ref{fig4} allows us to investigate the accuracy of the GC-isotropy and LSR-isotropy assumptions. We evaluated the simulated trajectories at $t=0,\,50,\,100,\,150$ and $200$ Myr, and considered the velocity vectors of the objects that are at these times in the solar neighbourhood (with speeds equal to $v$). These vectors are decomposed in order to obtain a $\vec{v}_t$ vector, which we combined with Eqs.\ \ref{eq3} and \ref{eq4} to determine the GC-isotropic and LSR-isotropic estimated distributions of $v_r$. We obtained the speeds estimated through assumption of isotropy ($v_{\text{iso}}$) by combining the $v_r$ estimates with the $\vec{v}_t$ vectors. The $v_{\text{iso}}/v$ distributions are then a proxy for the accuracy of the isotropy assumptions. In Fig.\ \ref{fig6} we show how the $v_{\text{iso}}/v$ distributions evolve over time, for the different Maxwellian kick distributions and for both GC-isotropy and LSR-isotropy. For $\sigma\geq100$ km/s the isotropies do not differ significantly in their accuracy, since high velocities are less affected by subtracting the LSR velocity. For low velocities ($\sigma\leq50$ km/s), however, the velocity vectors of the simulated objects are more or less aligned with $\vec{v}_{\text{LSR}}$, due to their initial circular velocity, because of which the GC-isotropic estimates underestimate the true speeds while the LSR-isotropic estimates remain remarkably accurate. This is why we are more confident in the LSR-isotropic velocity estimates shown in Fig.\ \ref{fig2}.
\section{Orbits}
\label{sec4}
Despite the galactocentric speeds of the BNSs harbouring little information about their kicks, we explored whether we can use the fact that the Galactic trajectories of the BNSs appear to be more or less circular \citep{Gaspari_2024} in order to constrain their systemic kicks. In order to do this, we determined the eccentricity of the Galactic orbits we simulated in Fig.\ \ref{fig4} and discuss the relationship between this eccentricity and kick distribution (Sect.\ \ref{sec4.1}). Moreover, we analysed the Monte Carlo estimated BNS trajectories and estimated their eccentricity (Sect.\ \ref{sec4.2}).
\begin{figure*}
    \centering
    \includegraphics[width=18cm]{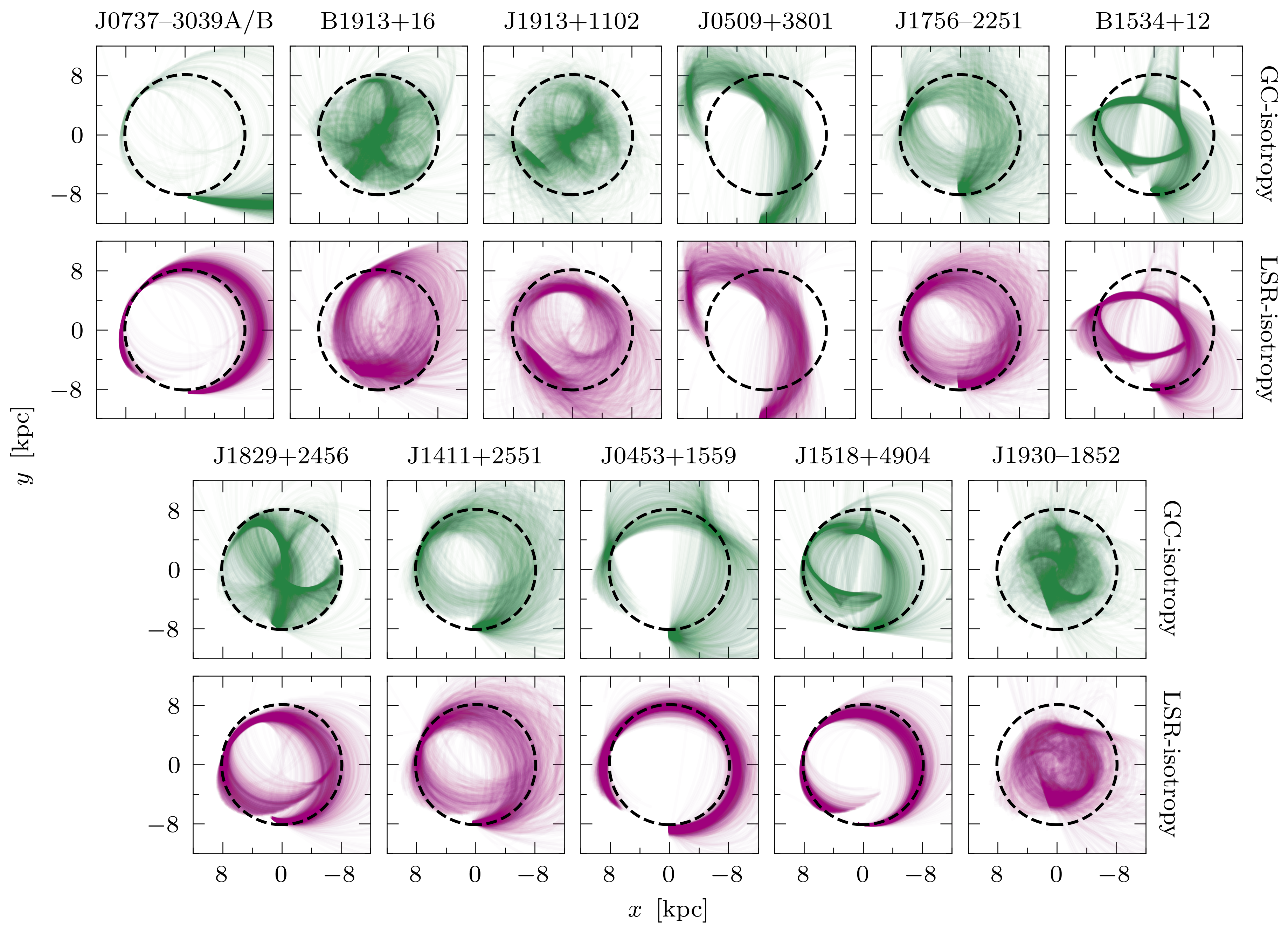}
    \caption{Kinematic histories of the Galactic BNSs (columns), assuming either GC-isotropy (green) or LSR-isotropy (purple) in the Monte Carlo estimation, integrated backwards in time for $200$ Myr. These trajectories are extended to $1$ Gyr when used in Sect.\ \ref{sec2.2} to determine the kinematic ages (and obtain $z$ values as shown in Appendix \ref{app.A}). The dashed circles show $R=R_{\sun}=8.122$ kpc \citep{Gravity_2018}.}
    \label{fig8}
\end{figure*}
\subsection{Eccentricity}
\label{sec4.1}
In Sect.\ \ref{sec3} we analysed the galactocentric velocities of the BNSs. However, when it comes to inferring kicks it may be more useful to look at their peculiar velocities (which are obtained by subtracting the local $\vec{v}_{\text{LSR}}$, as determined by the MW potential). After all, for small kicks the orbits of the objects remain more or less circular, which means they have a lower peculiar velocity ($v_{\text{pec}}$) than objects that receive high kicks, and retain these low velocities after $\sim40$ Myr. Nevertheless, for higher kicks $v_{\text{pec}}$ decreases similarly to $v$ \citep{Disberg_2024}, meaning it may still be difficult to differentiate between large kicks using the magnitude of $v_{\text{pec}}$. Thus, for objects older than $\sim40$ Myr the current magnitude of $v_{\text{pec}}$ is only useful for inferring kicks insofar as it can reveal how circular the Galactic trajectory of an object is.\\
\indent We therefore aim to investigate directly how the kick distribution affects the resulting Galactic orbits and their eccentricity, and whether we can use this to infer kicks. In particular, we are interested in the eccentricity of the orbits of the objects in Fig.\ \ref{fig4}, which received kicks from vastly different kick distributions. However, these Galactic orbits are not Keplerian, so they do not have an eccentricity as defined in Keplerian dynamics. Nevertheless, we define an analogue of eccentricity for Galactic orbits, by analysing the orbits of the kicked objects in our simulation and defining the minimum galactocentric radius an object obtains in their orbit as $R_{\min}$, and the maximum value of $R$ as $R_{\max}$. Then, we compute our eccentricity-analogue: 
\begin{equation}
    \label{eq6}
    \tilde{e}=\dfrac{R_{\max}-R_{\min}}{R_{\max}+R_{\min}}\quad,
\end{equation}
which we will refer to as the eccentricity of the Galactic orbit, or just the BNS' eccentricity (we note that this is not related to the eccentricity within the binary). Similarly to a Keplerian eccentricity: if $R_{\min}=R_{\max}$ the eccentricity equals $0$ and the orbit is perfectly circular, and if $R_{\min}=0$ or $R_{\max}\gg R_{\min}$ the eccentricity goes to $1$.\\
\begin{figure*}
    \centering
    \includegraphics[width=18cm]{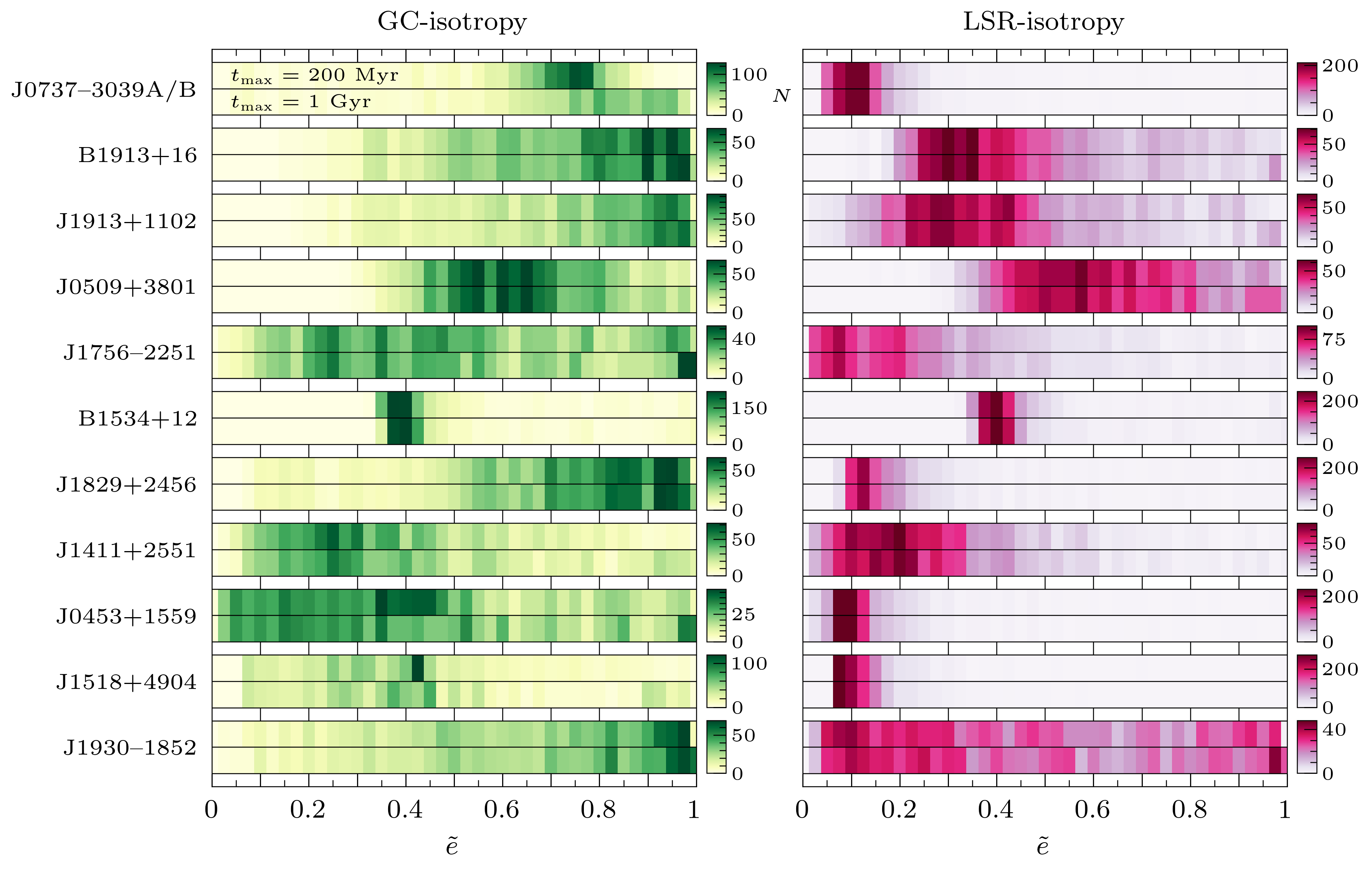}
    \caption{Eccentricities of the Galactic orbits of the BNSs, determined by applying Eq.\ \ref{eq6} to the Monte Carlo estimated orbits of each binary, assuming either GC-isotropy (left panel) or LSR-isotropy (right panel), and shown in 2D histograms with $\tilde{e}$ bins of $0.025$. For each BNS, we compute $\tilde{e}$ for the trajectories integrated back up to $t_{\max}=200$ Myr (as shown in Fig.\ \ref{fig8}), and also for the trajectories with $t_{\max}=1$ Gyr (as used in Sect.\ \ref{sec2.2}), and compare both $\tilde{e}$-distributions ($t_{\max}=200$ Myr first, and then $t_{\max}=1$ Gyr below).}
    \label{fig9}
\end{figure*}
\indent If we compute the Galactic orbits of kicked objects long enough, the values we find for $R_{\min}$ and $R_{\max}$ will correspond to the peri- and apocentres of the Galactic orbits, respectively. However, it is possible that within our simulations some objects do not have enough time to cross their peri- and apocentres, meaning $R_{\min}$ and $R_{\max}$ can depend -- to some degree -- on the computation time of the simulation ($t_{\max}$). Nevertheless, even if $t_{\max}$ does not allow for the simulated objects to cross their peri- and apocentres, our estimates of $\tilde{e}$ will still be an approximation of the Galactic eccentricity and tell us something about the circularity of their orbits. For example, if an object received a large kick, it might start moving radially outwards in such a way that $R_{\min}=R(t=0)$ and $R_{\max}=R(t=t_{\max})$, for a certain $t_{\max}$. Nevertheless, this will result in high values of $\tilde{e}$, even if this value will increase slightly if we increase $t_{\max}$.\\
\indent For the simulations shown in Fig.\ \ref{fig4} (where $t_{\max}=200$ Myr), we determined $\tilde{e}$ for each of the $10^3$ objects. In Fig.\ \ref{fig7}, we show for each simulation the eccentricities of the objects that are at a certain time in the solar neighbourhood. For each object, their value of $\tilde{e}$ does not change, but at each point in time there are different objects in the solar neighbourhood, resulting in different values of $\tilde{e}$ being observed. The figure shows that for Maxwellian kick distributions with increasing $\sigma$, we find objects in the solar neighbourhood with more eccentric orbits. That is, larger kicks perturb the initially circular orbits of the objects more, resulting in higher values of $\tilde{e}$. Moreover, the difference between $t\leq40$ Myr and $t>40$ Myr we observe in the galactocentric velocities of these objects (Fig.\ \ref{fig4}) is less prominent in their eccentricities. For $t\leq40$ Myr, the objects in the solar neighbourhood are slightly more eccentric, possibly due to objects receiving a significant kick because of which they do not return to the solar neighbourhood within $t_{\max}$. However, the median eccentricities do depend on $\sigma$, contrary to the velocities in Fig.\ \ref{fig4}, meaning $\tilde{e}$ can potentially allow us to distinguish between different kick distributions. Lastly, we note that in Fig.\ \ref{fig7}, similarly to Fig.\ \ref{fig4}, we find no significant differences between the results that use exponential disc and the ones that use the Gaussian annulus prior.
\subsection{Trajectories}
\label{sec4.2}
Since eccentricity can potentially provide insight in kicks, we investigated the eccentricities of the Galactic orbits of the BNSs. Similarly to Sect.\ \ref{sec2.2}, we traced the BNS trajectories back by estimating their present-day velocity vector and assuming either GC-isotropy or LSR-isotropy. In Fig.\ \ref{fig8}, we show these trajectories traced back for $200$ Myr (corresponding to $t_{\max}$ in Fig.\ \ref{fig7}). The figure shows that, despite the assumption of isotropy, most orbits are well-constrained, and some of these orbits (such as the LSR-isotropic orbits of J0737--3039A/B, J1829+2456, J1411+2551, J0453+1559, and J1518+4804) appear to have a remarkably low eccentricity, while others -- and in particular the GC-isotropic orbits -- seem to be more eccentric (such as the orbits of B1913+16, J0509+3801, B1534+12, and J1930--1852). However, not all orbits shown in Fig.\ \ref{fig8} show plausible kinematic histories of the BNSs. In particular, the GC-isotropic orbits of J0737--3039A/B mostly trace back the binary to large galactocentric radii, and do not return closer to the Galactic centre within $200$ Myr. Since J0737--3039B has a characteristic age of $\sim50$, and it is unlikely that this binary was formed this far away from the Galactic centre, we do not deem these particular trajectories an accurate representation of the actual kinematic history of this binary. This may suggest that, at least in this case, LSR-isotropic orbits represent a more physical scenario. Nevertheless, the other orbits appear to be plausible, allowing us to estimate the eccentricities of these Galactic orbits.\\
\begin{figure*}
    \centering
    \includegraphics[width=18cm]{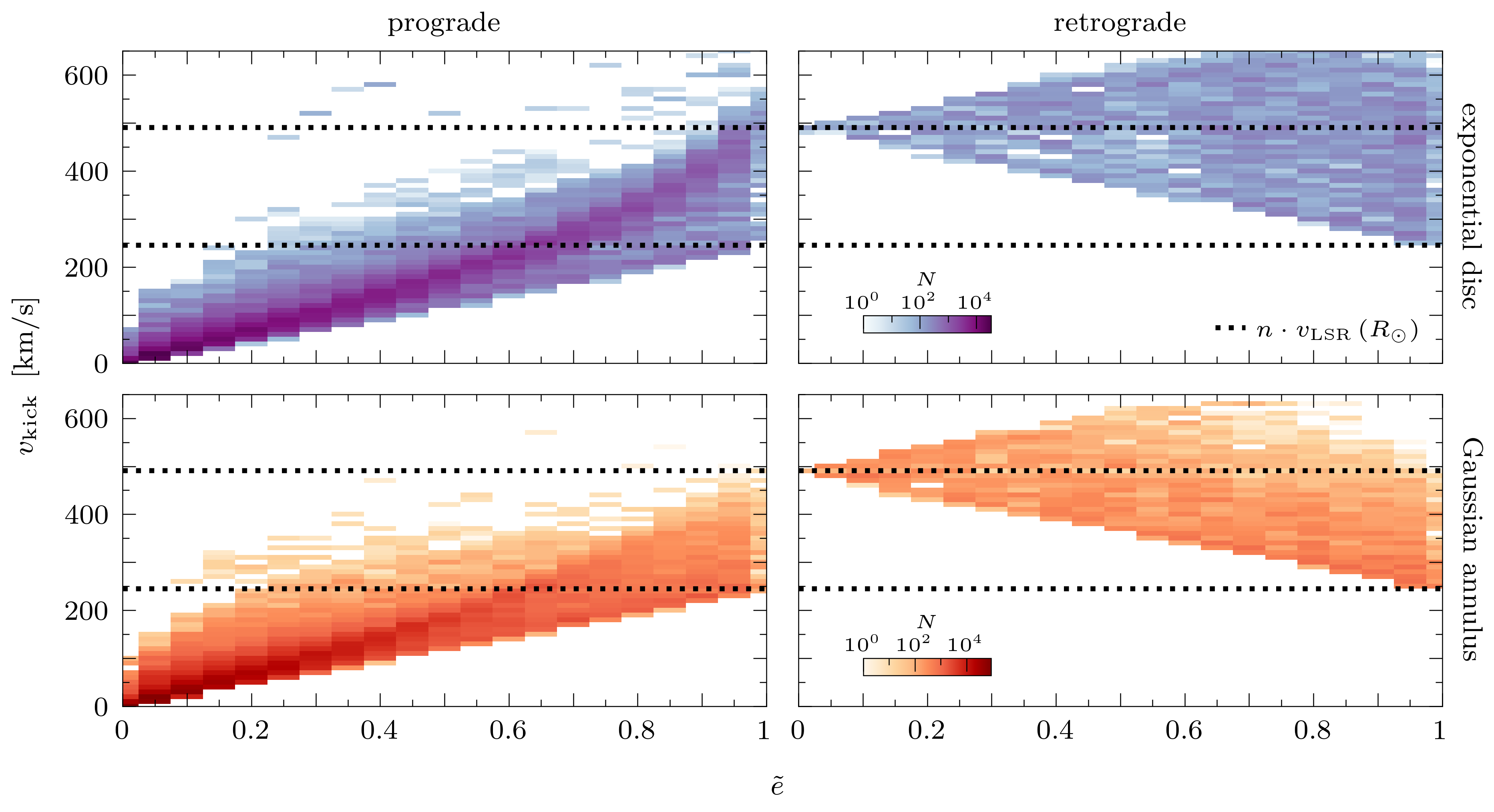}
    \caption{Number of eccentricities found in the solar neighbourhood for each $v_{\text{kick}}$, shown in a 2D histogram with $\tilde{e}$ bins of $0.05$ and $v_{\text{kick}}$ bins of $10$ km/s, on a logarithmic colour scale. For each $v_{\text{kick}}=n\cdot10$ km/s, we simulate $10^3$ objects using either the exponential disc (top row) or Gaussian annulus (bottom row) prior, and evaluate their orbits in between $40$ Myr and $t_{\max}=200$ Myr. The resulting eccentricities are divided into prograde (left column) and retrograde (right column) orbits. The dashed lines show $1$ and $2$ times the circular velocity ($v_{\text{LSR}}$) at a solar radius.}
    \label{fig10}
\end{figure*}
\indent We used the BNS trajectories to determine $R_{\min}$ and $R_{\max}$, and then calculated $\tilde{e}$ (through Eq.\ \ref{eq6}). This resulted in an $\tilde{e}$-distribution consisting of $10^3$ eccentricities, for each binary. We determined these distributions for $t_{\max}=200$ Myr and for $t_{\max}=1$ Gyr, and repeated this for both isotropy assumptions. Figure \ref{fig9} shows the resulting $\tilde{e}$-distributions. For the LSR-isotropic orbits, the eccentricity (1) is well-constrained, (2) obtains values where $\tilde{e}\lesssim0.5$ (with the exception of J0509+3801), and (3) does not change significantly between $t_{\max}=200$ Myr and $t_{\max}=1$ Gyr. This means that (1) the eccentricity distributions appear suitable for constraining the BNS kicks, (2) these kicks will probably be relatively small (considering the results from Fig.\ \ref{fig7}), and (3) within $200$ Myr most of these orbits cross their peri- and apocentres (making $\tilde{e}$ mostly independent of $t_{\max}$ if $t_{\max}\geq200$ Myr). For the GC-isotropic eccentricities in Fig.\ \ref{fig9}, the distributions are broader and obtain higher values of $\tilde{e}$, meaning that if we deduce kicks from these distributions the GC-isotropic kicks will be larger. In fact, for some binaries the LSR-isotropic values of $\tilde{e}$ are below $0.5$, while the GC-isotropic counterparts approach $1$ (e.g.\ B1913+16, J1913+1102, J1829+2456, and J1930--1852). Moreover, some of the GC-isotropic distributions shift -- to a certain degree -- to higher values between $t_{\max}=200$ Myr and $t_{\max}=1$ Gyr. After all, for low kicks the assumption of GC-isotropy underestimates $v$ and therefore overestimates $\tilde{e}$.\\
\indent In particular, the GC-isotropic eccentricities of J0737--3039A/B are dependent on $t_{\max}$. After all, these orbits are moving away from the Galactic centre (as shown in Fig.\ \ref{fig8}), meaning that increasing $t_{\max}$ will also increase $R_{\max}$ and therefore $\tilde{e}$ as well. The other GC-isotropic orbits show a similar effect, albeit less strong: there is a slight pile-up at $\tilde{e}\gtrsim0.8$ for $t_{\max}=1$ Gyr. Nevertheless, the distributions are mostly stable, due to most trajectories crossing their peri- and apocentres within $200$ Myr. The eccentricity of B1534+12 is particularly well-constrained, showing a narrow-peak at $\tilde{e}\approx0.4$, for both isotropies and both values of $t_{\max}$.
\section{Kicks}
\label{sec5}
Since we are interested in estimating the BNS kicks based on the eccentricities of their Galactic orbits, we investigated the relationship between kick and eccentricity (Sect.\ \ref{sec5.1}). Then, we applied this relationship to the eccentricity distributions we found for the Galactic BNSs in order to kinematically constrain their kicks (Sect.\ \ref{sec5.2}).
\subsection{Kicks versus eccentricity}
\label{sec5.1}
In Fig.\ \ref{fig7} we already found a relationship between kicks and eccentricities: higher kicks cause more eccentric orbits. In order to show the exact relationship between kick magnitude and eccentricity, we expanded this simulation. That is, we repeated our simulation (as described in Sect.\ \ref{sec3.1}), but instead of a Maxwellian kick distribution we used a delta function to describe the kicks, and repeated the simulation for values of $v_{\text{kick}}$ between $0$ km/s and $650$ km/s, with steps of $10$ km/s (and for both the exponential disc and Gaussian annulus prior). For these simulations we chose $t_{\max}=200$ Myr, since we found in Fig.\ \ref{fig9} that the $\tilde{e}$-values of the BNSs do not change significantly beyond $200$ Myr. Increasing $t_{\max}$ beyond this value would therefore not affect our results meaningfully. Then, we evaluated the trajectories of the simulated objects every $1$ Myr, and determined the eccentricities of the orbits that are in the solar neighbourhood at that point in time. Because young objects show slightly higher values of $\tilde{e}$ (as found in Fig.\ \ref{fig7}), we left out these objects and summed all eccentricities we find in between $40$ Myr and $t_{\max}$, effectively determining the relationship between $v_{\text{kick}}$ and $\tilde{e}$.\\
\begin{figure*}
    \centering
    \includegraphics[width=18cm]{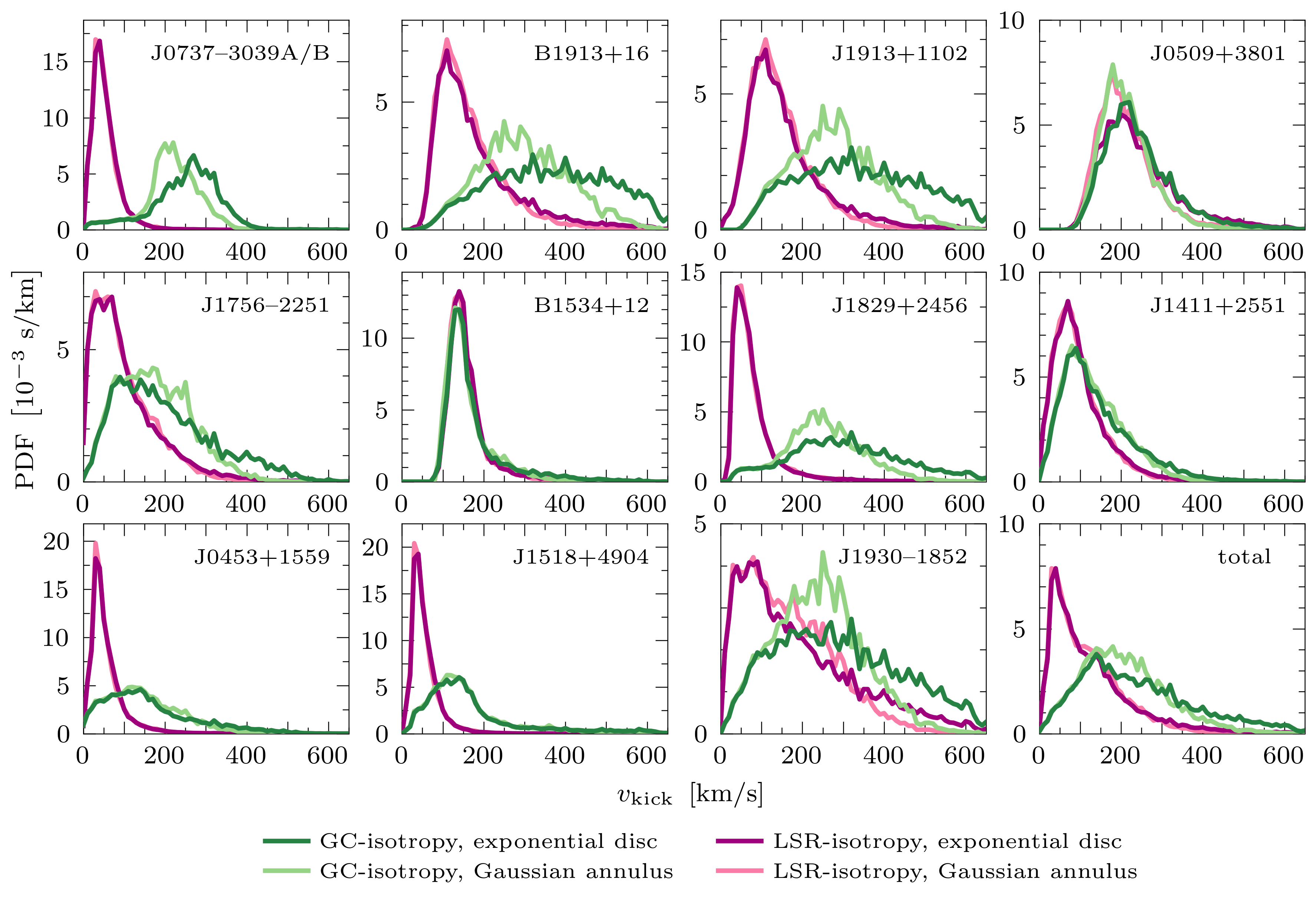}
    \caption{Posterior kick distributions resulting from our eccentricity-constrained estimates, for GC-isotropy (green lines) and LSR isotropy (purple lines), and for the exponential disc (dark lines) and Gaussian annulus (light lines) prior. The panels show the normalised results from integrating the posteriors in Appendix \ref{app.B} over $\tilde{e}$, meaning the lines trace the values of a histogram with bins of $10$ km/s. Each panel shows the distributions for an individual BNS, except for the lower right panel which combines all distributions.}
    \label{fig11}
\end{figure*}
\indent However, if we only consider the eccentricity of an orbit without taking its direction into account, this relationship will be bimodal. After all, if we observe an object with $\tilde{e}=0$ moving along with the Solar System, we estimate that $v_{\text{kick}}=0$ km/s: its initial circular orbit has not been disturbed. If, on the other hand, we observe an object with $\tilde{e}=0$ orbiting the Galactic centre in \emph{opposite} direction compared to the Solar System, this would mean that it received a large kick which reversed its initial circular orbit, coincidentally resulting in a circular orbit in opposite direction. This shows that in addition to $\tilde{e}$, we need to take into account whether the Galactic orbit of the objects are in the same or opposite direction of the Solar System. We call these two cases \textquotedblleft prograde\textquotedblright\ and \textquotedblleft retrograde\textquotedblright\ orbits, respectively. For the BNS trajectories in Fig.\ \ref{fig8} and the trajectories of the objects in our simulation, we know whether they are prograde or retrograde, since we can compare their initial velocity vector to the solar velocity, and if $v_{\phi}/v_{\phi,\sun}\geq0$ the entire orbit is prograde (due to conservation of angular momentum). For each BNS, $\gtrsim90\%$ of their LSR-isotropic trajectories are prograde. The GC-isotropic trajectories are $>90\%$ prograde, except for the GC-isotropic orbits of B1913+16 ($54\%$), J1913+1102 ($61\%$), J1829+2456 ($77\%$), and J1930--1852 ($73\%$).\\
\indent In Fig.\ \ref{fig10}, we show the results of these simulations, separated in prograde and retrograde orbits, for the exponential disc and the Gaussian annulus prior. If $v_{\text{kick}}=0$ km/s, all orbits are prograde with $\tilde{e}=0$: the objects retain their initial, circular orbit. Then, when we increase $v_{\text{kick}}$, the Galactic orbits become more eccentric. If $v_{\text{kick}}$ equals the circular velocity ($v_{\text{LSR}}$) at a solar radius, which is approximately $v_{\text{LSR}}$ of the solar neighbourhood, it becomes possible for the kick to be of equal magnitude but opposite direction compared to the initial velocity vector. This results in a net velocity of $0$ km/s, causing the object to fall to the Galactic centre, obtain $R_{\min}=0$ kpc and therefore $\tilde{e}=1$. If $v_{\text{kick}}>v_{\text{LSR}}\left(R_{\sun}\right)$, it is possible for the kick to be pointed in opposite direction of the initial circular velocity, and cause retrograde orbits in the solar neighbourhood. At this point, a larger kick means that the retrograde orbits can obtain velocities closer to $v_{\text{LSR}}\left(R_{\sun}\right)$, resulting in a decrease of the retrograde values of $\tilde{e}$. This continues up until $v_{\text{kick}}=2v_{\text{LSR}}\left(R_{\sun}\right)$, where the kick can cause perfectly circular, retrograde orbits. Increasing $v_{\text{kick}}$ beyond this point will result again in more eccentric orbits, approaching $\lim_{v_{\text{kick}}\to\infty}\tilde{e}=1$.\\
\indent We do not find significant differences between the results that use the exponential disc and the ones that use the Gaussian annulus prior, in particular for $\tilde{e}\lesssim0.6$ the distributions are remarkably similar. For $\tilde{e}\gtrsim0.6$, there are a few differences between the two rows in Fig.\ \ref{fig9}. The prograde orbits which use the Gaussian annulus prior and are extremely eccentric tend to be the results of slightly lower kicks than their exponential disc counterparts. This can be explained by the fact that the exponential disc prior seeds the objects closer the Galactic centre and thus deeper in the Galactic potential, meaning that it takes a larger kick for them to obtain orbits that travel to extremely large galactocentric radii and do not cross the solar neighbourhood between $40$ Myr and $200$ Myr. In other words, we expect that for highly eccentric orbits (i.e.\ $\tilde{e}\gtrsim0.8$), the results start to depend -- to some degree -- on $t_{\max}$. This can also be found in the retrograde orbits for the Gaussian annulus: $v_{\text{kick}}\gtrsim500$ km/s results in the most eccentric orbits to not be found in the solar neighbourhood (these kicks are also large enough to cause objects to become unbound, depending on the alignment between kick and initial velocity vector). However, for our purposes the relationship between $v_{\text{kick}}$ and $\tilde{e}$ found in Fig.\ \ref{fig10} suffices, since the trajectories of the BNSs mostly contain prograde orbits with $\tilde{e}<0.8$, except for a few GC-isotropic orbits (which we are less confident in).
\subsection{Kick velocities}
\label{sec5.2}
Now that we have estimated the distributions of $\tilde{e}$ for the BNSs (Fig.\ \ref{fig9}) and have determined the relationship between $v_{\text{kick}}$ and $\tilde{e}$ (Fig.\ \ref{fig10}), we combined the two in order to estimate the BNS kicks. That is, for the BNSs we took each value of $\tilde{e}$ in their eccentricity distribution, determined whether the corresponding orbit is prograde or retrograde, and matched the value to the corresponding column in Fig.\ \ref{fig10}. Then, we took out that column, normalised it, and put it in a posterior distribution. We repeated this (1) for each $\tilde{e}$ in the distribution of Fig.\ \ref{fig9}, (2) for the exponential disc and the Gaussian annulus prior, and (3) for the GC-istropy and LSR-isotropy assumption. Effectively, this means we weighted each column in Fig.\ \ref{fig10} based on the $\tilde{e}$ distributions we found in Fig.\ \ref{fig9} (and combined the prograde and retrograde orbits). In Appendix \ref{app.B} we show the posterior distributions for each binary. We integrated these posteriors over $\tilde{e}$ and normalised the result, which gave the posterior probability distributions of $v_{\text{kick}}$ (as discussed in Appendix \ref{app.B}).\\
\indent In Fig.\ \ref{fig11} we show our estimated distributions of $v_{\text{kick}}$ for each individual BNS, as well as the combined kick distribution. Due to the LSR-isotropic orbits being more circular, they result in smaller estimated kicks compared to the GC-isotropic ones. GC-isotropy, after all, tends to underestimate low velocities (Fig.\ \ref{fig6}), resulting in an overestimated eccentricity (Fig.\ \ref{fig9}), and therefore also overestimate the corresponding kick magnitude. Nevertheless, for some BNSs (i.e.\ J0509+3801, B1534+12 and J1411+2551) the LSR-istropic and GC-isotropic estimates agree well. Moreover, the exponential disc and Gaussian annulus prior result in similar kick estimates, with the GC-isotropic estimates for J0737--3039A/B, B1913+16, J1913+1102, and J1930--1852 showing the largest differences. This is caused by their highly eccentric orbits (Fig.\ \ref{fig9}) and the fact that the major differences between the two priors occur at high eccentricities (Fig.\ \ref{fig10}). Because of this, together with the fact that (1) some GC-isotropic orbits do not appear to be plausible kinematic histories of the BNSs, (2) the highly eccentric orbits in Fig.\ \ref{fig10} start to be dependent on $t_{\max}$, and (3) LSR-isotropic velocity estimates are more accurate for low velocities, we are more confident in the LSR-isotropic kick estimates.\\
\indent The lower right panel in Fig.\ \ref{fig11} contains the total kick distributions for the Galactic BNSs, showing that the LSR-isotropic estimates peak at $\sim40-50$ km/s and the GC-isotropic estimates peak at $\sim150-200$ km/s. In Appendix \ref{app.C} we fit a log-normal and a Maxwellian distribution to these estimates, and show that the LSR-isotropic estimates -- which we are more confident in than their GC-isotropic counterparts -- peak only slightly below the kick distribution found by \citet{O'Doherty_2023}, who estimate the kicks of NSs with low-mass (non-NS) companions using the method of \citet{Atri_2019}. We find that our LSR-isotropic estimates (i.e.\ the average between the results from the exponential disc and Gaussian annulus prior) are well-described by a log-normal distribution that peaks at $43\pm2$ km/s, and note that $v_{\text{kick}}\leq50$ km/s is usually seen as the region of low or no kicks, since it equals the typical escape velocity of globular clusters \citep[e.g.][]{Atri_2019,Mandel_2020,Willcox_2021,O'Doherty_2023}. In our log-normal fit to the LSR-isotropic estimates, $23\%$ of the kick velocities are $\leq50$ km/s.\\
\indent The method we used to constrain the BNS kicks is to some degree similar to the method \citet{Atri_2019} applied to black hole X-ray binaries. They Monte Carlo estimated the 3D velocity vector of these binaries and then traced their trajectories back in time. Then, instead of analysing the eccentricity of the entire orbit, they looked at the peculiar velocities at the instances when the binaries crossed the disc and considered these as potential kick velocities. Using this method, they obtain a potential kick distribution for each binary (\citeauthor{O'Doherty_2023} \citeyear{O'Doherty_2023} note that this method is slightly biased towards low kicks, since these produce more disc crossings, which can be corrected for by considering the same amount of disc crossings for each trajectory). As a comparison between our method and the method of \citet{Atri_2019}, we considered the first disc crossings of the BNS trajectories ($\chi_1$, used to estimate $\tau_{\text{kin}}$ in Sect.\ \ref{sec2.2} and displayed in Appendix \ref{app.A}), and determined their peculiar velocities at these points. We compare these peculiar velocities to our kick estimates in Appendix \ref{app.D}, and find that they match remarkably well. Our method does, however, differ from the method of \citet{Atri_2019}, meaning different assumptions were made in order to estimate the kicks. For example, our method assumes a prior distribution of objects before they are kicked (Fig.\ \ref{fig3}), whereas their method assumes that averaging peculiar velocities over disc crossings approximates the kick velocity. Also, we note that both methods assume that objects are initially located at $z=0$ with no peculiar velocity, but this can be potentially be altered in order to describe objects for which this is not an accurate description.
\section{Conclusion}
\label{sec6}
In this work, we determined the characteristic spin-down and kinematic ages of the Galactic BNSs (Sect.\ \ref{sec2}), and examined whether their galactocentric speeds are representative of their kicks (Sect.\ \ref{sec3}). As a novel method to constrain the BNS kicks, we determined the eccentricity of their Galactic orbits (Sect.\ \ref{sec4}) and estimated their kicks through these eccentricities (Sect.\ \ref{sec5}). Below, we summarise our findings for each section:
\begin{itemize}
    \item The characteristic ages (through Table \ref{tab1} and Eq.\ \ref{eq2}) and the kinematic ages \citep[through the velocity vectors estimated with the method of][which assumes either GC-isotropy or LSR-isotropy]{Gaspari_2024} of the Galactic BNSs indicate that they are likely to have ages $\gtrsim40$ Myr (Fig.\ \ref{fig1}), even though it remains difficult to relate these estimates to the true ages of the BNSs.
    \item The Galactic BNSs have median galactocentric speeds of $\sim200-250$ km/s (Fig.\ \ref{fig2}). However, using the findings of \citet{Disberg_2024}, and expanding them with a different prior spatial distribution (Fig.\ \ref{fig3}), we showed that vastly different kick distributions all lead to median galactocentric speeds in the solar neighbourhood of $\sim150-250$ km/s (Fig.\ \ref{fig4}) after $\sim40$ Myr (Fig.\ \ref{fig5}). Since the BNS ages likely exceed this value, this means we cannot constrain their kicks based on these speeds. Also, we find that GC-istotropic estimates tend to underestimate low velocities, while LSR-isotropic estimates remain accurate (Fig.\ \ref{fig6}). 
    \item However, we find that despite the fact that different kick distributions result in similar galactocentric speeds, the eccentricities (Eq.\ \ref{eq6}) of their Galactic trajectories do depend on the kick distribution, also for $t\gtrsim40$ Myr (Fig.\ \ref{fig7}). This indicates that if we investigate the Galactic orbits of the BNSs (Fig.\ \ref{fig8}), and determine their eccentricity (Fig.\ \ref{fig9}), we can potentially infer their kick velocities.
    \item In order to do this, we simulated populations of kicked objects and determine the eccentricities of the Galactic orbits of the objects that are in the solar neighbourhood between $40$ Myr and $200$ Myr. We repeated this for different kick values, which resulted in a relationship between kick and eccentricity (Fig.\ \ref{fig10}). Combining this relationship with the BNS eccentricities we found, we constructed kick estimates for each individual BNS, as well as a total kick distribution (Fig.\ \ref{fig11}). We find that the exponential disc and Gaussian annulus prior both give similar results, whereas the GC-isotropy results in higher kick estimates than the LSR-isotropy counterparts. We are more confident in the LSR-isotropy estimates, and find that they are well-described by a log-normal distribution peaking at $\sim40-50$ km/s.
\end{itemize}
Moreover, we note that (1) \citet{O'Doherty_2023} find a kick distribution for NSs with low-mass companions which peaks at slightly higher values compared to our (LSR-isotropic) distribution (Appendix \ref{app.C}), and (2) applying the method of \citet{Atri_2019} to the Galactic BNSs gives results similar to ours, even if we just limit this to the first disc crossing (Appendix \ref{app.D}).\\
\indent Our results are made uncertain by several factors. For instance, in our simulation we assume that when the BNSs are kicked, they are on perfectly circular Galactic orbits (i.e.\ they have no peculiar velocity) at $z=0$ kpc. This means that our results are uncertain to the degree that BNSs can have non-zero peculiar velocities pre-kick, and are located at $|z|>0$ kpc when they are kicked. It is likely that before the final kick at the time of the second supernova, the progenitors have already acquired peculiar velocities due to the previous SN event. Moreover, the BNS kick distribution we find is a description of eleven observed Galactic BNSs. Because of this, it is not obvious that the distribution we find is representative of all BNS kicks. It is, for example, conceivable that BNS merger times depend on the magnitude of their kick \citep[e.g.][]{Beniamini_2024}, possibly resulting in a bias against high kicks, or are perhaps influenced by their Galactic trajectories \citep{Stegmann_2024}. One could also suspect that highly kicked objects are less likely to be observed close to the Solar System. However, we argue that Fig.\ \ref{fig10} captures kicks up to at least $500$ km/s: the figure shows that objects that received kicks in between our Galactic BNS kick estimates and $\gtrsim500$ km/s do cross the solar neighbourhood within the time frame of the simulations. Since these kicks do not seem to be present in our BNS sample, albeit a relatively small sample, we argue that our simulation accounts for this bias up to $v_{\text{kick}}\approx500$ km/s. Nevertheless, we cannot decisively exclude the possibility of an undetected BNS population that receives extremely high kicks ($\gtrsim500$ km/s). We do note that these uncertainties apply to our method as well as the method of \citet{Atri_2019}.\\
\indent \citet{Behroozi_2014} find that in order to explain the observed offsets of SGRBs, one needs a SGRB progenitor kick distribution where one in five objects receives a kick with a magnitude $>150$ km/s. In the log-normal fit to our (LSR-isotropic) results, the region where $v_{\text{kick}}>150$ km/s contains approximately $30\%$ of the probability density, which is more or less consistent with the percentage required by \citet{Behroozi_2014}. We therefore state that, while the BNS kick distribution we find may peak at low kicks ($\sim40-50$ km/s), it still contains the amount of high kicks required by \citet{Behroozi_2014} to match the galaxy-offset distribution found by \citet{Fong_2013}.\\
\indent Future research could expand on several aspects of our analysis, but it would be particularly interesting to expand the simulations behind Fig.\ \ref{fig10}. That is, it would be interesting to increase (1) the size of the simulations, (2) the value of $t_{\max}$, and (3) the range of kicks being sampled, in order to estimate the relationship between kick and the eccentricity of the Galactic orbit more robustly. This could give a more accurate estimate for the kick corresponding to a highly eccentric orbit ($\tilde{e}\gtrsim0.9$), after which it could be applied to objects that are thought to receive higher kicks (such as isolated pulsars). After all, our estimation does not use the fact that the objects we describe are BNSs: it can potentially be applied to all kicked objects, in order to shed more light on kicks and kick dynamics. Moreover, it would be interesting to investigate how the fact that BNS kicks consist of two separate kicks -- due to the two SNe in the binary -- affects our findings. These kicks, in turn, could possibly affect the merger times or radio lifetimes of the binaries, which may provide additional constraints on the estimated BNS kicks. This way, our general method for inferring kicks can be tailored for neutron star binaries.

\begin{acknowledgements}
    We thank Fiorenzo Stoppa for useful discussions regarding this Master's project, as well as the referee for comments that helped to improve this paper. AJL was supported by the European Research Council (ERC) under the European Union’s Horizon 2020 research and innovation programme (grant agreement No. 725246), and NG acknowledges studentship support from the Dutch Research Council (NWO) under the project number 680.92.18.02. In this work, we made use of \lstinline{NUMPY} \citep{Harris_2020}, \lstinline{SCIPY} \citep{Virtanen_2020}, \lstinline{MATPLOTLIB} \citep{Hunter_2007}, \lstinline{GALPY} \citep{Bovy_2015}, and \lstinline{ASTROPY}, a community-developed core Python package and an ecosystem of tools and resources for astronomy \citep{Astropy_2013,Astropy_2018,Astropy_2022}.
\end{acknowledgements}
\bibliographystyle{TeXnical/aa}
\bibliography{References}

\begin{appendix}
\section{Disc crossings}
\label{app.A}
\begin{figure*}
    \centering
    \includegraphics[width=18cm]{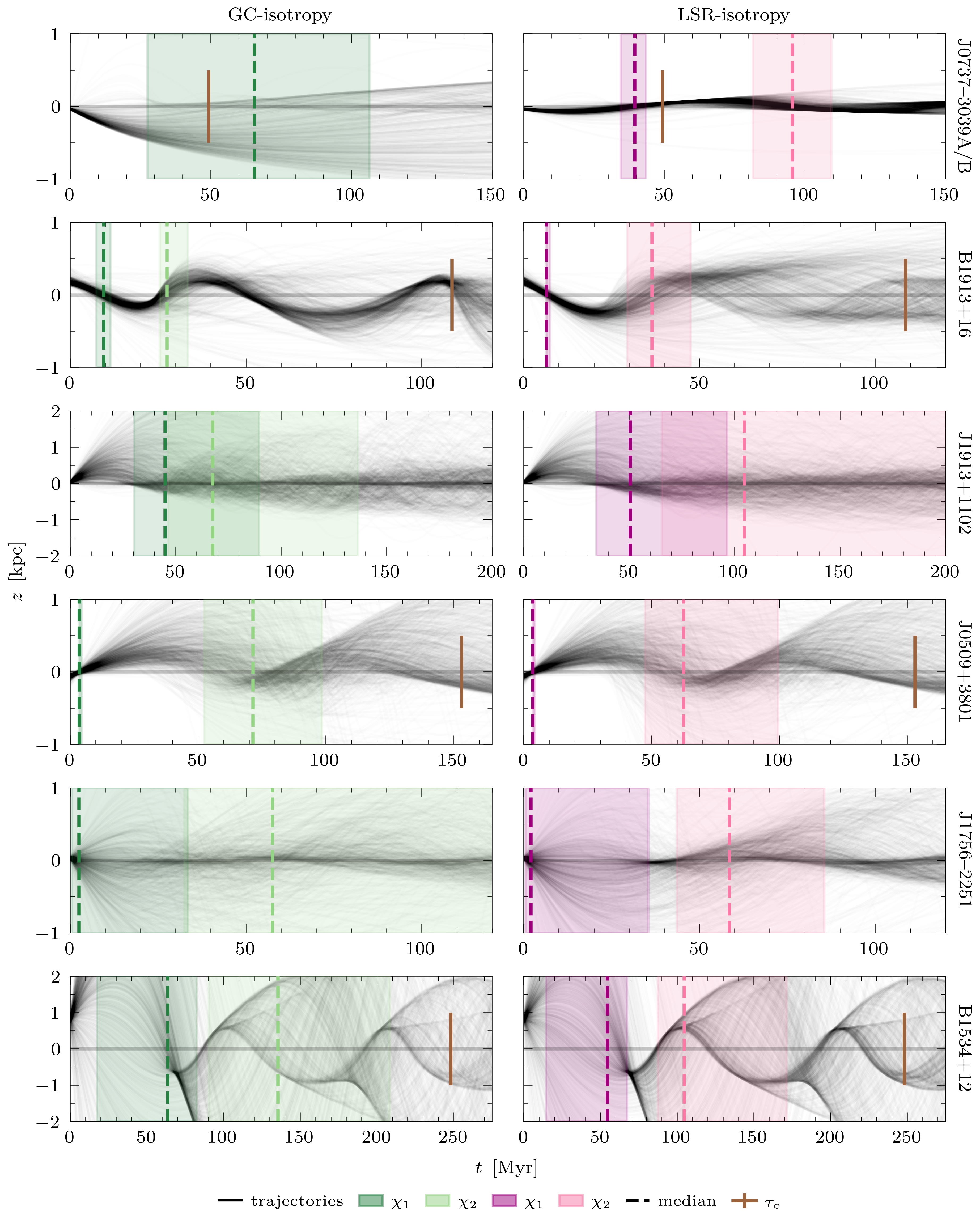}
    \caption{Disc-crossings of the first six Galactic BNSs, estimated using the method described in Sect.\ \ref{sec2.2}, for GC-isotropy (left column, through Eq.\ \ref{eq3}) and LSR-isotropy (right column, through Eq.\ \ref{eq4}). The black lines show the overplotted trajectories of the BNSs, where the dark green and dark purple regions correspond to the $68\%$ intervals of $\chi_1$, and the light green and light purple regions correspond to the $68\%$ intervals for $\chi_2$ (as given in Table \ref{tab3}). The dashed lines of the same colour indicate the medians in these regions. The brown plusses show $\tau_{\text{c}}$, where the horizontal line extends to the $68\%$ interval and the vertical line is located at the median (which are also given in Table \ref{tab3}).}
    \label{figA1}
\end{figure*}
\begin{figure*}
    \centering
    \includegraphics[width=18cm]{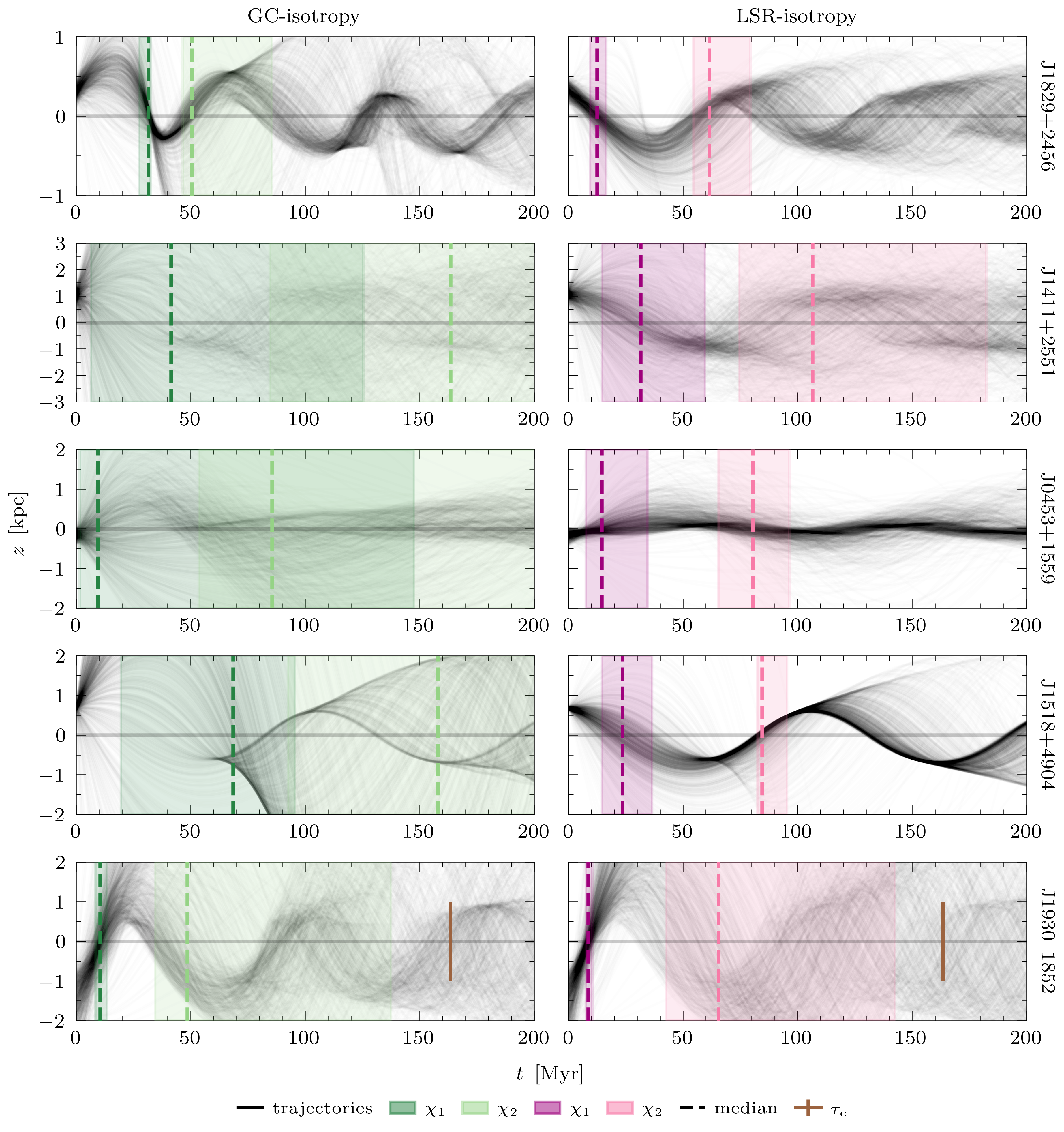}
    \caption{Disc-crossings of the latter five Galactic BNSs, similarly to Fig.\ \ref{figA1}, estimated using the method described in Sect.\ \ref{sec2.2}, for GC-isotropy (left column, through Eq.\ \ref{eq3}) and LSR-isotropy (right column, through Eq.\ \ref{eq4}). The black lines show the overplotted trajectories of the BNSs, where the dark green and dark purple regions correspond to the $68\%$ intervals of $\chi_1$, and the light green and light purple regions correspond to the $68\%$ intervals for $\chi_2$ (as given in Table \ref{tab3}). The dashed lines of the same colour indicate the medians in these regions. The brown plusses show $\tau_{\text{c}}$, where the horizontal line extends to the $68\%$ interval and the vertical line is located at the median.}
    \label{figA2}
\end{figure*}
In Sect.\ \ref{sec2}, we discuss different quantities that can tell us something about the ages of the Galactic BNSs (i.e. characteristic ages and kinematic ages), and show the estimated distributions in Fig.\ \ref{fig1} and Table \ref{tab3}. The kinematic ages ($\tau_{\text{kin}}$) shown in the table are determined through a Monte Carlo estimation of the BNS orbits, assuming either GC-isotropy or LSR-isotropy (Sect.\ \ref{sec2.2}). Fig.\ \ref{fig1} shows the distributions of $\tau_{\text{kin}}$, but in Figs.\ \ref{figA1} and \ref{figA2} we also display the trajectories of the BNS orbits in the $z$ dimension, in order to show how the estimated $\tau_{\text{kin}}$ and their $68\%$ intervals correspond to the actual trajectories of the BNSs and their disc crossings. For some binaries, and mostly in the case of LSR-isotropy, the orbits are well-constrained and $\tau_{\text{kin}}$ is a good estimate of the disc-crossings. For others, the trajectories are less constrained by the isotropy assumptions, resulting in a larger uncertainty on $\tau_{\text{kin}}$ (on top of the methodological uncertainties in these kinematic ages). In Fig.\ \ref{figA1}, the $\tau_{\text{kin}}\approx5$ Myr solution for J0747--3039A/B by \citet{Willems_2006} can be seen in the GC-isotropic case.
\section{Posteriors}
\label{app.B}
\begin{figure*}
    \centering
    \includegraphics[width=18cm]{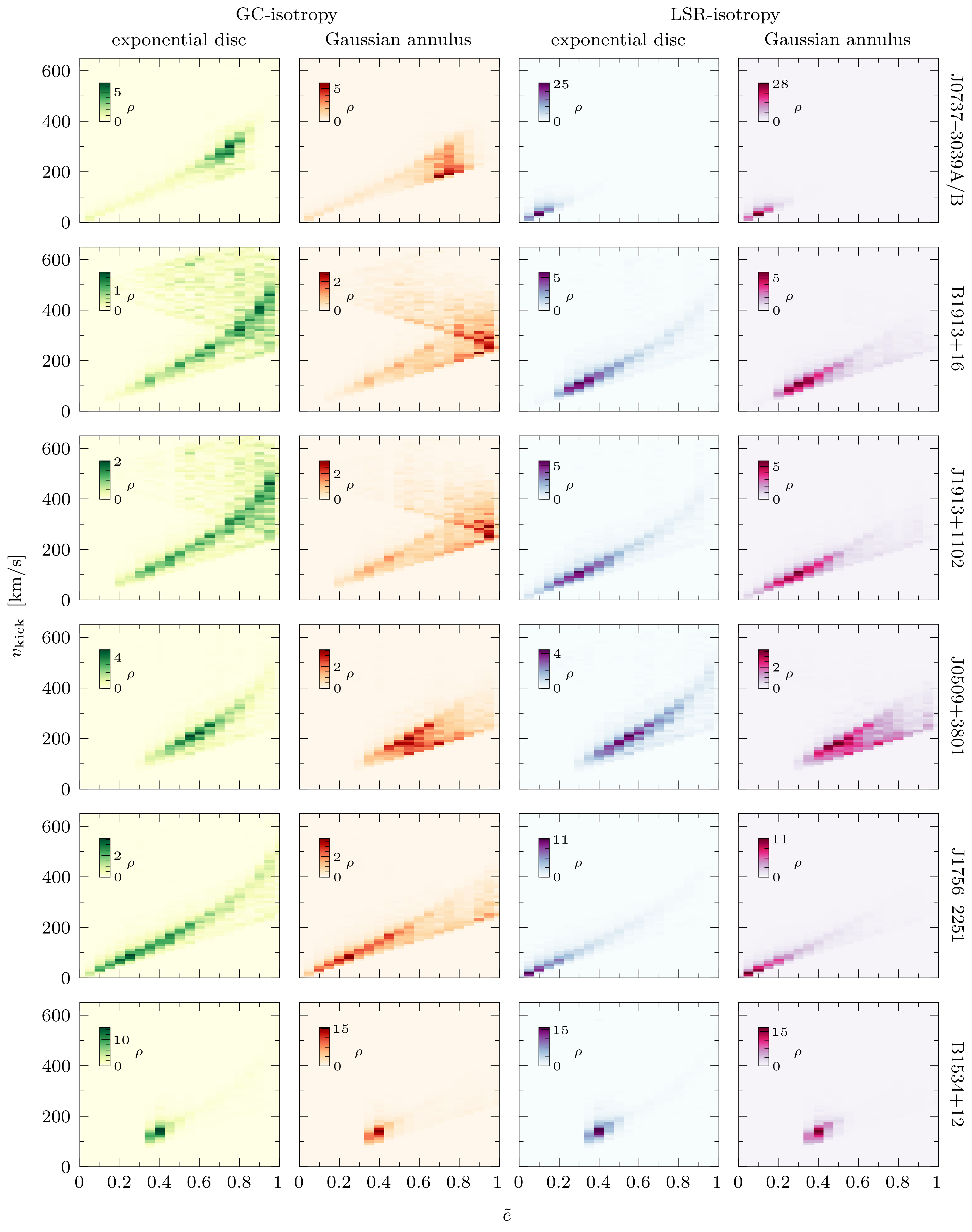}
    \caption{Posterior distributions of $v_{\text{kick}}$ versus $\tilde{e}$, for the first six BNSs (rows), following either the GC-isotropic (two left columns) or LSR-isotropic (two right columns) estimations of $\tilde{e}$, as shown in Fig.\ \ref{fig9}. For each isotropy assumption, we compute the posterior using the results from the exponential disc (left column) or Gaussian annulus (right column) prior, as shown in Fig.\ \ref{fig10}. The posteriors are shown in 2D histograms with $\tilde{e}$ bins of $0.05$ and $v_{\text{kick}}$ bins of $10$ km/s, and are normalised so that the (linear) colour scale shows the probability density ($\rho$) in units of $10^{-2}$ s/km.}
    \label{figB1}
\end{figure*}
\begin{figure*}
    \centering
    \includegraphics[width=18cm]{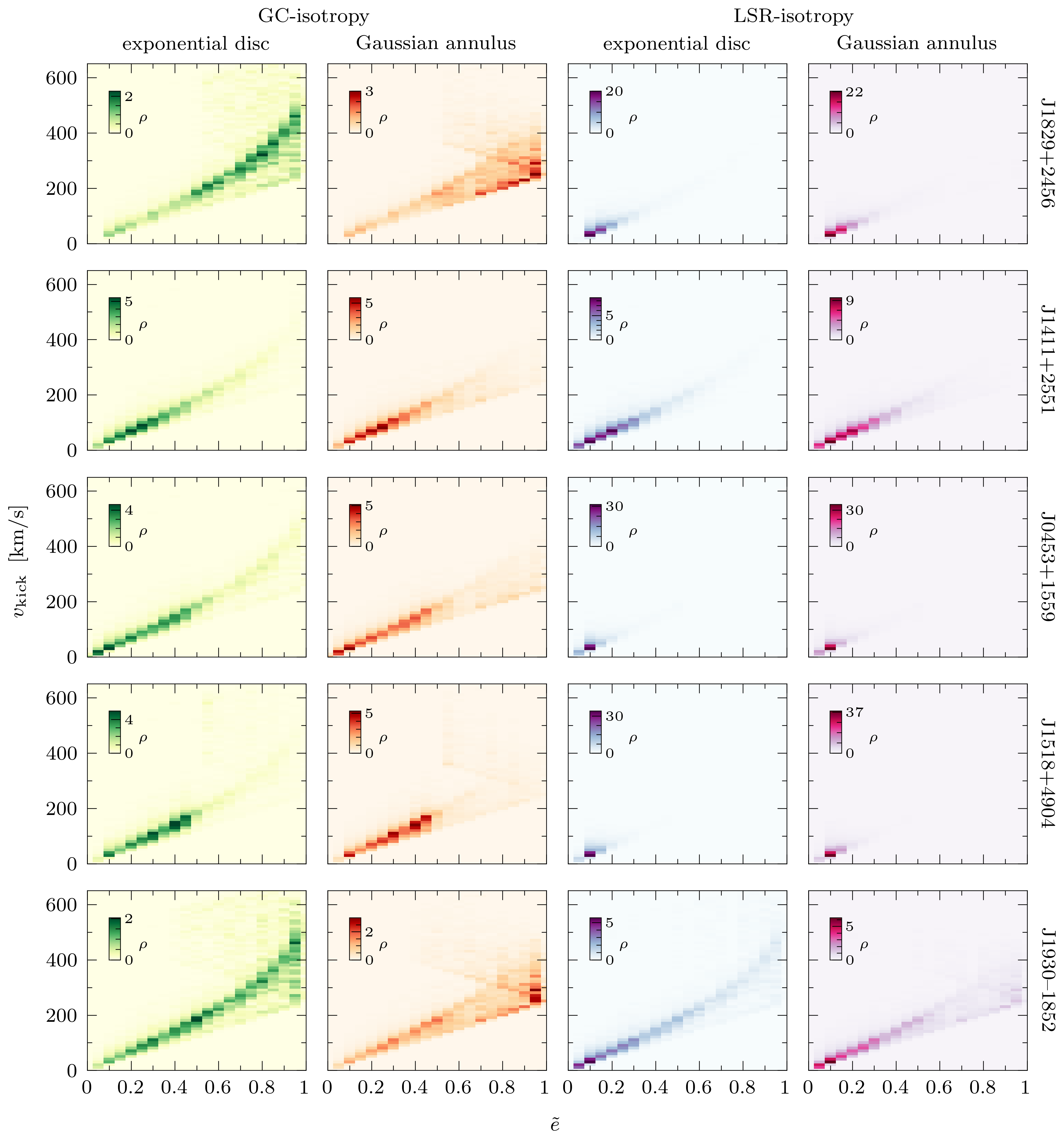}
    \caption{Posterior distributions of $v_{\text{kick}}$ versus $\tilde{e}$ (similar to \ref{figB1}), for the latter five BNSs (rows), following either the GC-isotropic (two left columns) or LSR-isotropic (two right columns) estimations of $\tilde{e}$, as shown in Fig.\ \ref{fig9}. For each isotropy assumption, we compute the posterior using the results from the exponential disc (left column) or Gaussian annulus (right column) prior, as shown in Fig.\ \ref{fig10}. The posteriors are shown in 2D histograms with $\tilde{e}$ bins of $0.05$ and $v_{\text{kick}}$ bins of $10$ km/s, and are normalised so that the (linear) colour scale shows the probability density ($\rho$) in units of $10^{-2}$ s/km.}
    \label{figB2}
\end{figure*}
In Sect.\ \ref{sec5}, we describe how we combine our estimates of the BNSs' eccentricities (Fig.\ \ref{fig9}) and the relationship we find between $v_{\text{kick}}$ and $\tilde{e}$ (Fig.\ \ref{fig10}), to determine the posterior kick estimates (Fig.\ \ref{fig11}). In order to infer $v_{\text{kick}}$ based on $\tilde{e}$, this method employs Bayes' theorem:
\begin{equation}
    \label{eqb1}
    P(v_{\text{kick}}\hspace{.5mm}|\hspace{.5mm}\tilde{e})=\dfrac{P(\tilde{e}\hspace{.5mm}|\hspace{.5mm}v_{\text{kick}})P(v_{\text{kick}})}{P(\tilde{e})}\quad.
\end{equation}
Since we want to relate this theorem to the results of our simulations as shown in Fig.\ \ref{fig10}, we define each cell in this figure to have a value equal to $N(\tilde{e}\hspace{.5mm}|\hspace{.5mm}v_{\text{kick}})$ and a size equals to $\Delta\tilde{e}\cdot\Delta v_{\text{kick}}$, while the amount of objects in each column and row equal $N(\tilde{e})$ and $N(v_{\text{kick}})$, respectively. Using these quantities, we determine that $P(\tilde{e}\hspace{.5mm}|\hspace{.5mm}v_{\text{kick}})=N(\tilde{e}\hspace{.5mm}|\hspace{.5mm}v_{\text{kick}})/N(v_{\text{kick}})$. Then, we define the prior $P(v_{\text{kick}})$ as proportional to the size of each row (i.e.\ $P(v_{\text{kick}})\propto N(v_{\text{kick}})$), effectively accounting for the fact that the fraction of objects that cross the solar neighbourhood differ for each $v_{\text{kick}}$, independent of $\tilde{e}$. These two relations allow us to determine $P(\tilde{e})$:
\begin{equation}
    \label{eqb2}
    P(\tilde{e})=\int P(\tilde{e}\hspace{.5mm}|\hspace{.5mm}v_{\text{kick}})P(v_{\text{kick}})dv_{\text{kick}}\propto\sum_{v_{\text{kick}}} N(\tilde{e}\hspace{.5mm}|\hspace{.5mm}v_{\text{kick}})=N(\tilde{e})\quad.
\end{equation}
Combining these probabilities with Eq.\ \ref{eqb1} gives, then:
\begin{equation}
    \label{eqb3}
    P(v_{\text{kick}}\hspace{.5mm}|\hspace{.5mm}\tilde{e})\propto\dfrac{N(v_{\text{kick}}\hspace{.5mm}|\hspace{.5mm}\tilde{e})}{N(v_{\text{kick}})}\dfrac{N(v_{\text{kick}})}{N(\tilde{e})}=\dfrac{N(\tilde{e}\hspace{.5mm}|\hspace{.5mm}v_{\text{kick}})}{N(\tilde{e})}\quad,
\end{equation}
which corresponds to Fig.\ \ref{fig10} with each column normalised individually. This relation allows us to estimate kicks based on an observed eccentricity. However, we do not have a single value of $\tilde{e}$ but a distribution ($D_{\tilde{e}}$), which is why we combine Eq.\ \ref{eqb3} with the eccentricity distribution $P(\tilde{e}\hspace{.5mm}|\hspace{.5mm}D_{\tilde{e}})$, which are displayed in Fig.\ \ref{fig9}, in order to determine the desired kick estimate posteriors from Fig.\ \ref{fig11}:
\begin{equation}
    \label{eqb4}
    P(v_{\text{kick}}\hspace{.5mm}|\hspace{.5mm}D_{\tilde{e}})=\int P(v_{\text{kick}}\hspace{.5mm}|\hspace{.5mm}\tilde{e})P(\tilde{e}\hspace{.5mm}|\hspace{.5mm}D_{\tilde{e}})d\tilde{e}\quad,
\end{equation}
for which we show the product $P(v_{\text{kick}}\hspace{.5mm}|\hspace{.5mm}\tilde{e})P(\tilde{e}\hspace{.5mm}|\hspace{.5mm}D_{\tilde{e}})$ in Figs.\ \ref{figB1} and \ref{figB2}. By collecting normalised columns from Fig.\ \ref{fig10}, corresponding to the distributions in Fig.\ \ref{fig9}, integrating the posterior over $\tilde{e}$ and normalising the result, we effectively compute the kick posteriors following Eqs.\ \ref{eqb3} and \ref{eqb4}.
\section{Fits}
\label{app.C}
\begin{figure*}
    \centering
    \includegraphics[width=18cm]{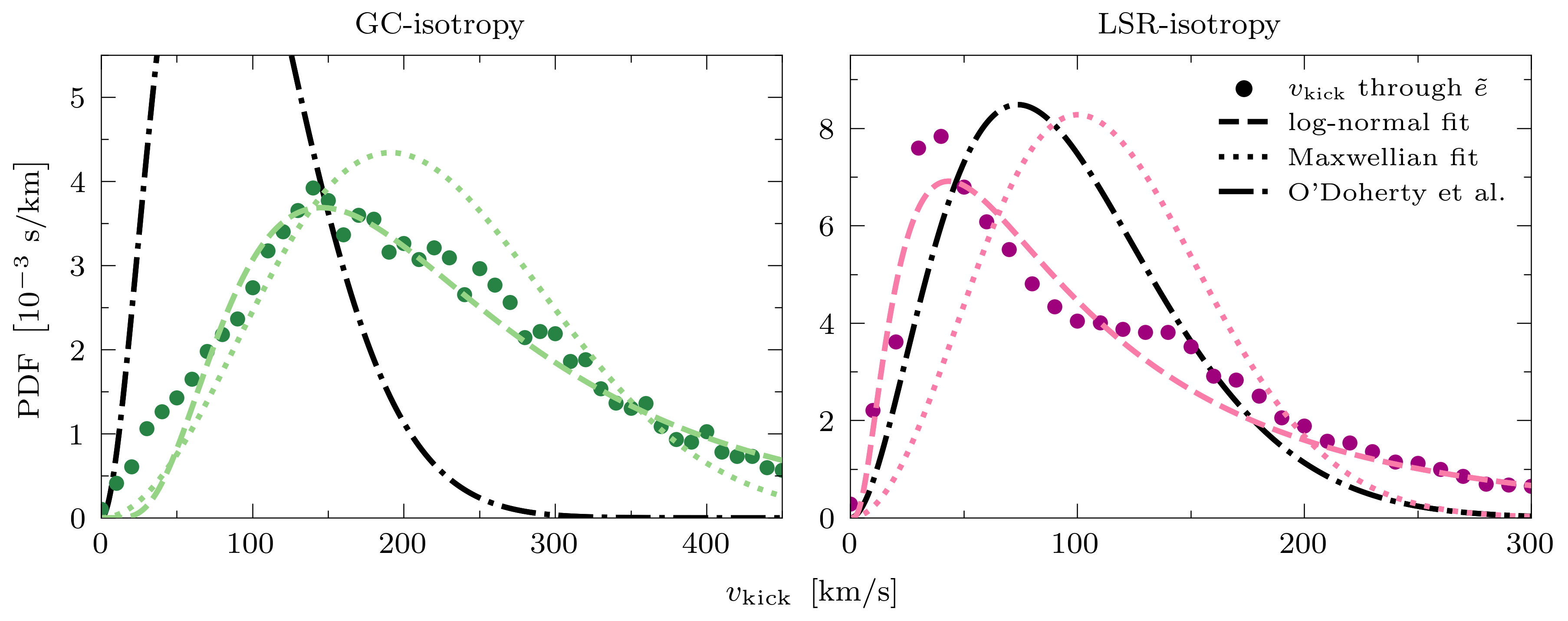}
    \caption{Total kick distribution as determined through $\tilde{e}$ (dots, average of the results using the exponential disc prior and the ones using the Gaussian annulus prior, tracing the tops of normalised histograms with bins of $10$ km/s), together with the fitted log-normal (dashed lines) and Maxwellian (dotted lines) distributions, for GC-isotropy (left panel) and LSR-isotropy (right panel). The fitted parameters of the distributions are given in Table \ref{tabc}. We also show the kick distribution that \citet{O'Doherty_2023} find for NSs with low-mass companions (black dash-dotted lines).}
    \label{figC}
\end{figure*}
In Fig.\ \ref{fig11} we show our kinematically constrained kick estimates for the individual BNSs, as well as a total distribution for all BNSs. To the total BNSs kicks, we fit a log-normal distribution: 
\begin{equation}
    \label{eqc1}
    \rho(x\hspace{.5mm}|\hspace{.5mm}\mu,\sigma)=\dfrac{1}{x\sigma\sqrt{2\pi}}\exp\left(-\dfrac{\left(\ln(x)-\mu\right)^2}{2\sigma^2}\right)\quad,
\end{equation}
and a Maxwellian:
\begin{equation}
    \label{eqc2}
    \rho(x\hspace{.5mm}|\hspace{.5mm}\sigma)=\sqrt{\dfrac{2}{\pi}}\dfrac{x^2}{\sigma^3}\exp\left(-\dfrac{x^2}{2\sigma^2}\right)\quad.
\end{equation}
We use a non-linear least-squares fit and obtain the optimal parameter values given in Table \ref{tabc}. In Fig.\ \ref{figC}, we show the fitted distributions, as well as the results from \citet{O'Doherty_2023}, who estimate the kicks of NSs with low-mass companions, and describe this with a beta distribution. The figure shows that the log-normal distribution seems to fit our results significantly better than the Maxwellians.
\begin{table}[h]
\centering
\caption{Parameters of the distributions fitted to our results, for a log-normal distribution (Eq.\ \ref{eqc1}) and a Maxwellian distribution (Eq.\ \ref{eqc2}).}
\label{tabc}
\begin{tabular}{lcc}
\hline
\hline\\[-10pt]
Log-normal            & GC-isotropy & LSR-isotropy \\ \hline\\[-10pt]
$\mu$                 & $5.36(1)$   & $4.57(2)$    \\
$\sigma$              & $0.62(1)$   & $0.89(2)$    \\
mode\tablefootmark{a} & $146(3)$    & $43(2)$      \\
                      &             &              \\ \hline\\[-10pt]
Maxwellian            &             &              \\ \hline\\[-10pt]
$\sigma$              & $135(2)$    & $71(3)$      \\
mode\tablefootmark{b} & $191(3)$    & $100(4)$     
\end{tabular}
\tablefoot{
Values in brackets are the $68\%$ uncertainty on preceding digits.\\
\tablefoottext{a}{Equals $\exp\left(\mu-\sigma^2\right)$.}
\tablefoottext{b}{Equals $\sqrt{2}\sigma$.}
}
\end{table}
\section{Peculiar velocities}
\label{app.D}
\begin{figure*}
    \centering
    \includegraphics[width=18cm]{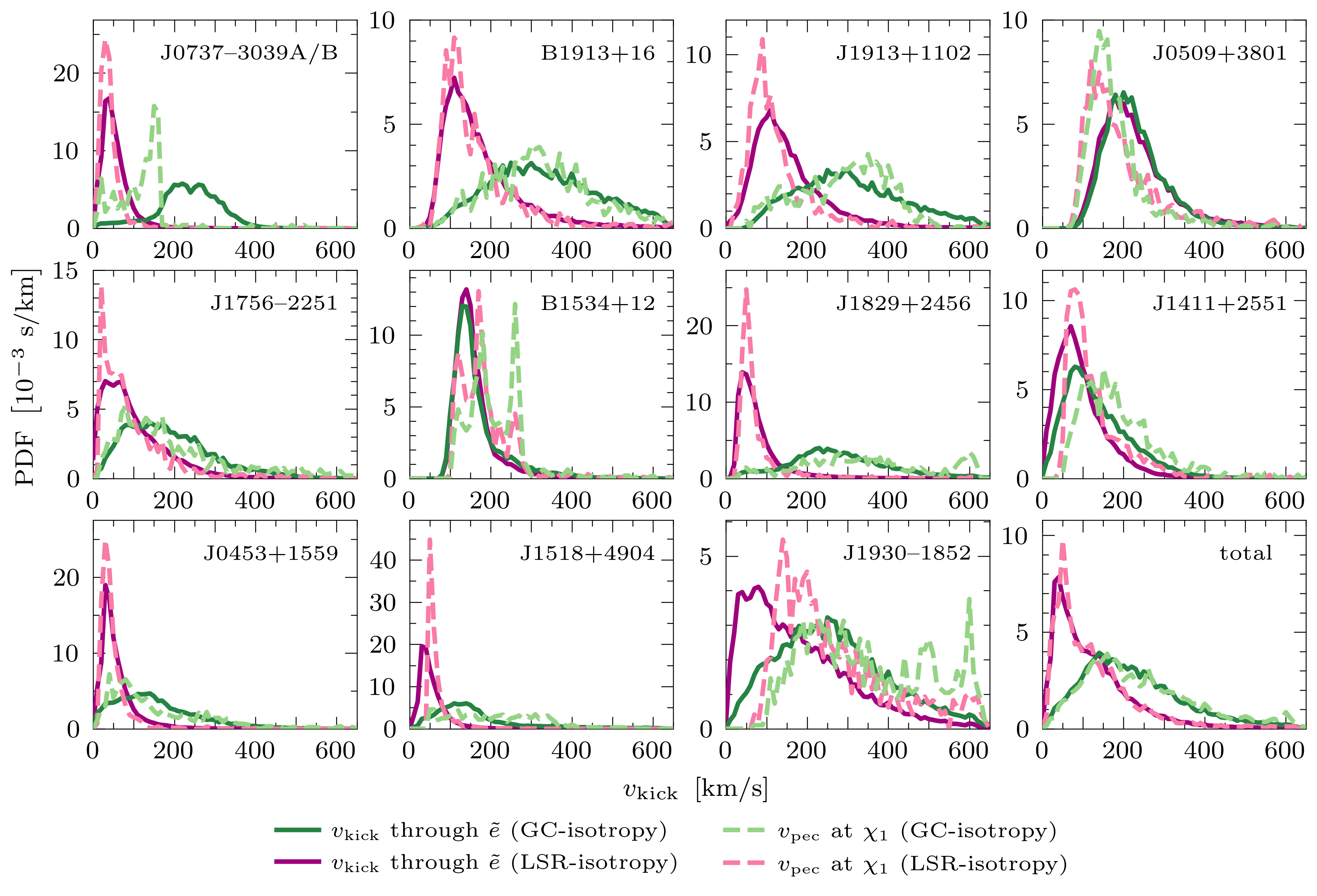}
    \caption{Comparison between our kick estimates as constrained through $\tilde{e}$ (solid lines) and peculiar velocities ($v_{\text{pec}}$) at the first disc crossing (dashed lines), for GC-isotropy (green lines) and LSR-isotropy (purple lines). For both isotropies, we show the average between the results using the exponential disc prior and the ones using the Gaussian annulus prior (as shown in Fig.\ \ref{fig11}). The lines trace normalised histograms with bins of $10$ km/s.}
    \label{figD}
\end{figure*}
As a comparison between our method and the method proposed by \citet{Atri_2019}, we compare our kick estimates to the peculiar velocities at the first disc crossings of the BNSs. For each BNS, we take the disc crossings that we used in order to estimate $\tau_{\text{kin}}$ for $\chi_1$ (in Sect.\ \ref{sec2.2}), and determine the peculiar velocity at that point as a potential kick velocity. Now, \citet{Atri_2019} do not limit themselves to the first disc crossing, but trace back the trajectories of the black hole X-ray binaries in their sample for $10$ Gyr and analyse the disc crossings during this entire period. Nevertheless, we find that even if we limit ourselves to the first disc crossing, our kick estimates match the peculiar velocities at $\chi_1$ well (as we show in Fig.\ \ref{figD}). In particular, the total kick distribution for the BNS sample looks nearly indistinguishable if determined through peculiar velocities at $\chi_1$, compared to our estimates through $\tilde{e}$. The most notable differences in the kick estimates occur for J0737--3039A/B, J0509+3801, B1534+12, and J1930--1852.
\end{appendix}

\end{document}